\documentclass[sigconf]{acmart}

\usepackage[ruled,linesnumbered]{algorithm2e}
\usepackage{subfigure}
\usepackage{multirow}
\usepackage{multicol}
\usepackage{enumitem}

\AtBeginDocument{%
  \providecommand\BibTeX{{%
    \normalfont B\kern-0.5em{\scshape i\kern-0.25em b}\kern-0.8em\TeX}}}

\copyrightyear{2022}
\acmYear{2022}
\setcopyright{acmcopyright}
\acmConference[KDD '22] {Proceedings of the 28th ACM SIGKDD Conference on Knowledge Discovery and Data Mining}{August 14--18, 2022}{Washington, DC, USA.}
\acmBooktitle{Proceedings of the 28th ACM SIGKDD Conference on Knowledge Discovery and Data Mining (KDD '22), August 14--18, 2022, Washington, DC, USA}
\acmPrice{15.00}
\acmISBN{978-1-4503-9385-0/22/08}
\acmDOI{10.1145/3534678.3539480}
\begin{document}

% \title{Transformer-based Multi-Agent Actor Critic Method for Active Voltage Control}
\title{Stabilizing Voltage in Power Distribution Networks via Multi-Agent Reinforcement Learning with Transformer}

\author{Minrui Wang}
\authornote{Both authors contributed equally to this research.}
\email{wangminrui0804@mail.ustc.edu.cn}
\author{Mingxiao Feng}
\authornotemark[1]
\email{fmxustc@mail.ustc.edu.cn}
\affiliation{%
  \institution{University of Science and Technology of China}
  \city{Hefei}
  \state{Anhui}
  \country{China}
  \postcode{230026}
}

\author{Wengang Zhou}
\authornote{Corresponding authors: Wengang Zhou and Houqiang Li.}
\email{zhwg@ustc.edu.cn}
\affiliation{%
  \institution{University of Science and Technology of China}
  \city{Hefei}
  \state{Anhui}
  \country{China}
  \postcode{230026}
}
\affiliation{%
  \institution{Institute of Artificial Intelligence, Hefei Comprehensive National Science Center}
  \city{Hefei}
  \state{Anhui}
  \country{China}
}

\author{Houqiang Li}
\authornotemark[2]
\email{lihq@ustc.edu.cn}
\affiliation{%
  \institution{University of Science and Technology of China}
  \city{Hefei}
  \state{Anhui}
  \country{China}
  \postcode{230026}
}
\affiliation{%
  \institution{Institute of Artificial Intelligence, Hefei Comprehensive National Science Center}
  \city{Hefei}
  \state{Anhui}
  \country{China}
}

\renewcommand{\shortauthors}{Wang and Feng, et al.}

\begin{abstract}
The increased integration of renewable energy poses a slew of technical challenges for the operation of power distribution networks. Among them, voltage fluctuations caused by the instability of renewable energy are receiving increasing attention.
Utilizing MARL algorithms to coordinate multiple control units in the grid, which is able to handle rapid changes of power systems, has been widely studied in active voltage control task recently.
%However, existing approaches apply methods proposed in the MARL community to the power network directly, ignoring the unique nature of the grid.
However, existing approaches based on MARL ignore the unique nature of the grid and achieve limited performance.
In this paper, we introduce the transformer architecture to extract representations adapting to power network problems and propose a \textbf{T}ransformer-based \textbf{M}ulti-\textbf{A}gent \textbf{A}ctor-\textbf{C}ritic framework (T-MAAC) to stabilize voltage in power distribution networks. 
%improve the performance of state-of-the-art MARL algorithms in power systems.
In addition, we adopt a novel auxiliary-task training process tailored to the voltage control task, which improves the sample efficiency and facilitates the representation learning of the transformer-based model.
We couple T-MAAC with different multi-agent actor-critic algorithms, and the consistent improvements on the active voltage control task demonstrate the effectiveness of the proposed method.\footnote{Code will be released at \url{https://github.com/cjdjr/T-MAAC}.} 
\end{abstract}





\maketitle

\section{Introduction}
The development and utilization of renewable energy is critical for addressing current energy and environmental concerns. In recent years, distributed generations (DGs), \emph{e.g.}, roof-top photovoltaics (PVs), have been steadily connected to power distribution networks because of its particular environmental friendliness, economy and flexibility. 
However, the increasing penetration of PVs in the distribution network may cause voltage swing due to their rapid active power changes. 
The voltage fluctuation can be alleviated by utilizing the control flexibility provided by PV inverters and other controllable devices such as static var compensators (SVCs)\cite{AVC_online}. Therefore, an elaborate scheme is required to coordinate the control between these distributed devices based on local information to ensure stable operation of the entire power system, which is called active voltage control\cite{MAPDN}. 

% Various approaches\cite{AVC_attention,AVC_MAAVC,AVC_MATD3,AVC_online,AVC_SAC} that applied MARL to active voltage control problem have been proposed.
There have been some previous efforts dedicated to active voltage control, which can be classified into three broad groups from the perspective of the control framework: centralized, distributed autonomous, and distributed cooperative control\cite{AVC_MATD3}.
% The centralized control utilizes optimization algorithms to solve a centralized voltage regulation problem based on information of the whole power system, such as the Optimal Power Flow (OPF)\cite{OPF} and the stochasitc programming (SP)-based approaches\cite{SP}. However, these centralized methods suffer from heavy computational burden, especially in large-scale power systems.
% As for another branch of research, distributed autonomous control strategies are only performed based on local observations which is easy to implement but difficult to achieve global optimal solution due to the lack of cooperation between various control units.
As a promising solution, the distributed cooperative control enables collaboration between distinct control units via limited communication links. 
Among them, some approaches\cite{AVC_attention,AVC_MAAVC,AVC_MATD3,AVC_online,AVC_SAC} apply multi-agent reinforcement learning (MARL) to active voltage control. 
These MARL algorithms are based on the centralized training and decentralized execution framework, which extract knowledge from historical data and simulation environment.
The learned strategies can be deployed to the grid and achieve cooperative control without any communication devices.
These attempts of applying MARL to power network tasks have attracted a lot of attention because of their strong adaptability to the unknown dynamic in real-time.

\begin{figure}%[h]
  \centering
  \includegraphics[width=0.7\linewidth]{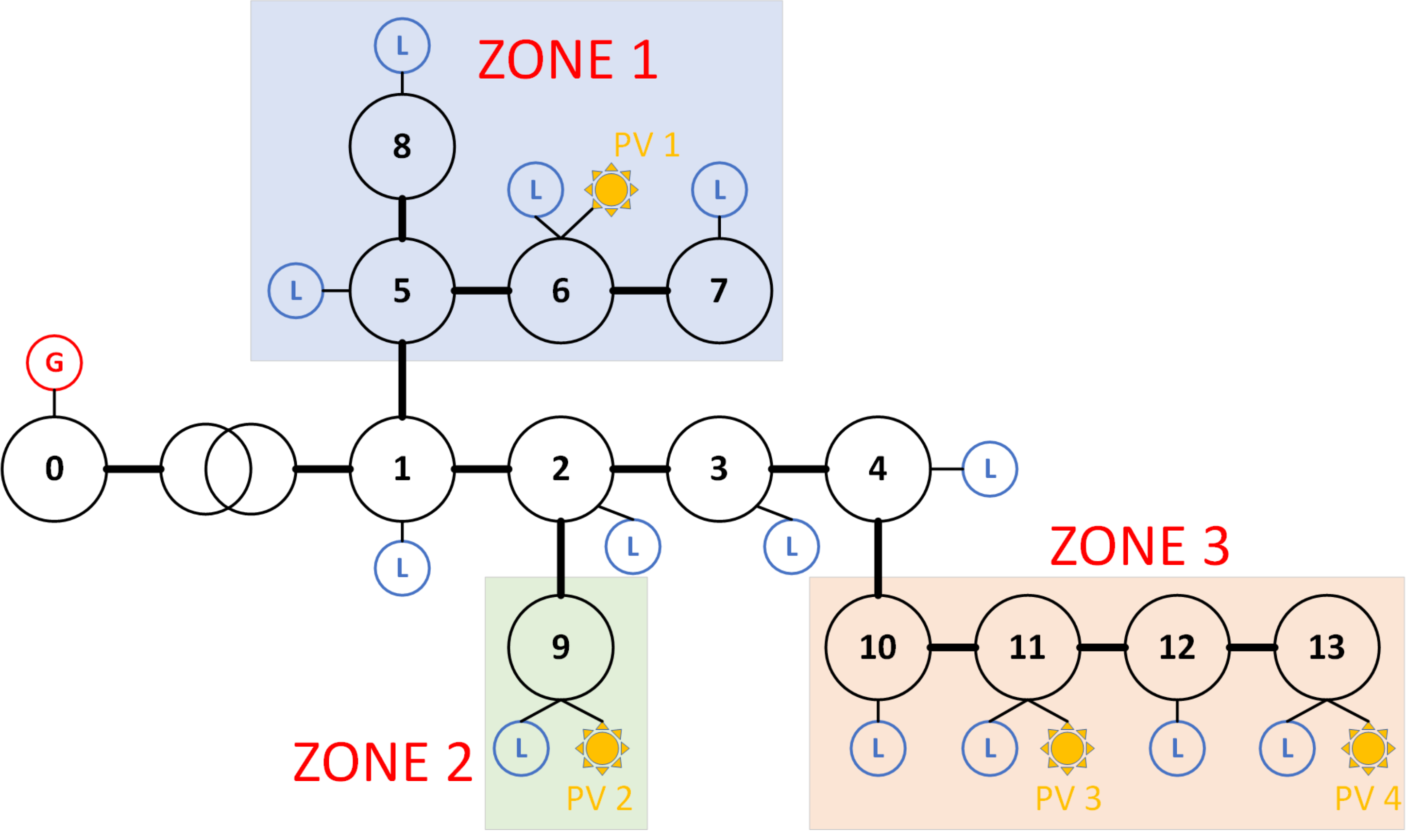}
  \caption{An example in power distribution network. Each bus in the distribution network is considered as a node in the graph. The whole network are divided into 4 zones connected to the main branch (node 1-4). We need to control voltages on node 1-13 and node 0 represent the main grid with constant voltage. $\mathbf{G}$ denotes an external generator; $\mathbf{L}$ denotes load connected to the node; and the sun emoji denotes the location a PV installed in.}
  \label{fig:illustration}
\end{figure}

For active voltage control problem, distributed control units are treated as agents, observing information about nodes in a zone of the grid. 
Figure \ref{fig:illustration} shows an example, in which the whole grid is divided into three zones and the sun emoji represents the location where a PV is installed.
Each PV inverter installed in a PV is regarded as a agent and observes information of nodes in the corresponding zone.
In prior works\cite{AVC_attention,AVC_MAAVC,AVC_MATD3,AVC_online,AVC_SAC}, the MLP-based policy networks and MLP-based critic network are applied to parameterize the policy functions and action-value functions, respectively\cite{Cooperative,MADDPG,MATD3,MAPPO,COMA}.
Besides, additional paddings are appended to each observation to guarantee that all observations have the same dimension. After that, the padded observations are mapped to control actions via policy networks whose parameters are shared among all agents.

It is worth noting that directly applying the above routine settings in the MARL community to the active voltage control task encounters a number of challenges:
\begin{enumerate}[leftmargin=*, topsep=0pt,parsep=0pt]
\item \textbf{Inconsistent number of nodes observed by various agents.} For example, as shown in Figure \ref{fig:illustration}, there are 4 nodes in zone 1 and 3, but only 1 node in zone 2.
\item \textbf{Inconsistent topology of nodes in each observation.} In Figure \ref{fig:illustration}, both zone 1 and zone 3 include 4 nodes, however they are connected in quite different ways.
\item \textbf{Inconsistent importance of nodes in a zone.} For example, typically, node 6 installed with PV has more frequent voltage fluctuation than other nodes in zone 1, which means that the PV 1 must take more attention to node 6 when making decisions.
\end{enumerate}
Current methods ignore these challenges and simply concatenate nodes information directly to obtain observations, with an assumption that neural networks are capable of automatically modeling the relationship between nodes in a sub-grid and decoupling observations smartly to address above challenges.
By following these settings, these approaches handle all information received from different agents in a same way and treat nodes in a observation uniformly with no regard of the topology of these nodes in the zone, which leads to limited representation learning capability.

To address the above challenges, we propose a transformer~\cite{transformer} based multi agent actor-critic framework (abbreviation as T-MAAC) for active voltage control, which can couple with mainstream multi-agent actor-critic algorithms.
Specifically, we propose a transformer-based policy network and a transformer-based critic network to obtain discriminative representations.
For the policy network, we divide the whole observation into node-based entities and project these node-based entities to high-dimensional semantic space via a transformer encoder. Inspired by~\cite{SMAAC,stabilize_transformer}, an adjacency matrix representing the connectivity of nodes in the zone is treated as the mask in the self-attention mechanism, introducing topological information to enhance representations. Then, we propose an embedding aggregation module to aggregate node-based information in the zone into the embedding from the agent's (control unit) point of view. To address the instability of transformer architecture in MARL algorithms\cite{updet, stabilize_transformer}, we develop a novel self-supervised auxiliary task in the training process of policy networks.
For the critic network, we exploit the vanilla transformer layer to approximate the global Q-value function. We introduce the self-attention mechanism to model the correlations between agents from the scale of the entire grid.
% To evaluate the performance of our proposed T-MAAC, we couple T-MAAC with two multi-agent actor-critic algorithms: MADDPG\cite{MADDPG} and MATD3\cite{MATD3}. The experimental results show that T-MAAC consistently improves the performance of these algorithms on 141-bus network and 322-bus network with three different reward functions based on MAPDN environment{\cite{MAPDN}}.
Experiments on the MAPDN environment\cite{MAPDN} demonstrate the effectiveness of our approach.

In summary, our main contributions are three-fold as follows:
\begin{itemize}[leftmargin=*,topsep=0pt,parsep=0pt]
% \item We analyze the challenges when applying policy networks and critic networks based on MLP to the active voltage control task.
\item We propose a novel transformer-based multi-agent actor-critic framework for active voltage control task and improve the performances of the existing multi-agent actor-critic algorithms from the perspective of voltage regulation and energy loss.
\item We adopt a self-supervised auxiliary task to stabilize the training process of MARL algorithms with transformer, improving the sample efficiency and facilitating the representation learning.
\item We introduce the attention mechanism into the voltage control in power network tasks, assisting a control unit in elaborating cooperative control strategies with other control units. It is more explainable and facilitates the MARL-based methods deploying to the realistic power system.
\end{itemize}

The rest of this paper is organized as follows. We first give a literature review on the related work in Section \ref{sec:related_work}. Then the background of active voltage control and the formulation of Markov Games are elaborated in Section \ref{sec:background}. The methodology is described in detail in Section \ref{sec:method}. Simulation results and discussions are presented in Section \ref{sec:experiments}. Finally, we conclude our work in Section \ref{sec:conclusions}.

\section{Related Works}
\label{sec:related_work}

\subsection{Multi-Agent Reinforcement Learning for Active Voltage Control}
Advances in machine learning lead to widespread applications of multi-agent reinforcement learning techniques to tackle active voltage control problem.
In \cite{AVC_MATD3}, authors take advantage of spectral clustering algorithms to partition the large distribution power network into several zones and formulate the control between each zones as Markov Game solved by MATD3.
\cite{AVC_attention} introduced an attention mechanism in the critic network to enhance scalability of algorithms. In contrast to \cite{AVC_attention}, we not only introduce attention mechanisms to model the relationship of agents in the critic network, but also propose transformer-based architecture adapting to the grid topology in the policy network. \cite{AVC_MAAVC} developed a approach with a manually designed voltage inner loop for the autonomous voltage control of transmission network based on MADDPG. \cite{AVC_SAC} leverages the sparse pseudo-Gaussian process to build a surrogate model using few-shot recorded data and control actions based on multi-agent soft actor critic algorithms.
Above prior works divided the whole grid into several zones, with each zone controlled by a single agent.
However, MAPDN\cite{MAPDN} modeled the active voltage control problem as a Dec-POMDP\cite{Dec-POMDP} and each PV inverter is controlled by an agent.
In this paper, we also follow the settings of Dec-POMDP proposed in MAPDN, which enables the presence of multiple agents with similar observations in a zone.

\subsection{Attention Mechanisms and Transformer in Multi-Agent Reinforcement Learning}
Transformers\cite{transformer} have been applied successfully to solve a wide variety of tasks in natural language processing\cite{bert} and computer vision\cite{detr,vit}.
However, transformers have not yet been fully explored in multi-agent reinforcement learning, mostly due to differing nature of problem, such as high variance in training.
UPDet\cite{updet} proposed the transformer-based individual value function and policy decoupling strategy based on value-based MARL algorithms to achieve improvements on multi-agent games in the StarCraft II environment with discrete action space.
Hierarchical RNNS-Based Transformers MADDPG (HRTMADDPG)\cite{hierarchical_transformer} combined transformers with RNNs, capturing the causal relationship between sub-time sequences.
\cite{self-attention-based} proposed a self-attention-based multi-agent continuous control algorithms to solve the problem of uneven learning degree and improved learning efficiency when faced with more agents.
The above works showed promising performance on game-like environments with a few agents.
However, it is not yet clear whether multi-agent algorithms with transformer can still achieve competitive performance if applied to the power system with large-scale agents, i.e. there are 38 agents on 322-bus network in MAPDN environment\cite{MAPDN}.
\section{Problem Formulation}
\label{sec:background}
\subsection{Active Voltage Control on Power Distribution Networks}
% In recent years, distributed generations (DG), e.g. roof-top photovoltaics (PVs), are gradually being connected to power distribution networks for their unique environmental friendliness, economy and flexibility.
% In a traditional radial distribution network, the electricity is generated from power plant and consumed by industrial and commercial customers, while the voltage in the grid decreases gradually with feeders.
% However, power flow can be injected into the grid due to the installation of DGs, and the voltage rises at the power injection point instead. Conventional methods to solve voltage fluctuation problem are not able to cope with various situations. A complex and elaborate scheme needs to be designed to coordinate these DGs and regulate the voltage of whole power distribution networks with limited local information, which is called active voltage control\cite{MAPDN, AVC1, AVC2, AVC3, AVC4}.

In this paper, a power distribution network installed with roof-top photovoltaics (PVs) is modeled as a tree graph structure $\mathcal{G}=(V,E)$ shown in Figure \ref{fig:illustration}, where $V = \{ 0,1,...,N \}$ and $E = \{ 1,2,...,N \}$ denote the set of nodes (buses) and edges (branches), respectively\cite{OPF}.
For bus $i \in V$, let $v_i$ and $\theta_i$ be the magnitude and phase angle of complex voltage and $s_i = p_i + j q_i$ denotes the complex power injection. There are complex and non-linear relationships between these physical quantities that satisfy the power system dynamics rules\cite{MAPDN}. In particular, node 0 is connected to the main grid, which serves to balance the active and reactive power in the distribution network.
Nodes in the distribution network are divided into several zones based on their shortest path from the terminal to the main branch\cite{MAPDN}. Also, loads (e.g. residential and industrial clients) and PVs are connected to some of nodes. Each PV is equipped with an PV inverter that generates reactive power to control the voltage around the standard value denoted as $V_{ref}$.
For safe operation of the distribution network, 5\% voltage fluctuation is usually allowed. Let $v_0 = 1.0$ per unit $(p.u.)$, the voltage amplitude of each bus must satisfy the following inequality condition: $0.95 p.u. \leq v_i \leq 1.05 p.u. , \forall i \in V \setminus \{ 0 \} $.
In the middle of day, the solar energy is converted into electrical energy and injected into the distribution network via PVs, which would increase $v_i$ out of the safe range. In contrast, $v_i$ may drop below the $0.95 p.u.$ due to the heavy load at night.
% To cope with these complex and challenging scenarios, we propose a multi-agent reinforcement learning (MARL) based approach to learn the collaborative control strategies among these PVs.
In this paper, we consider each PV inverter installed in a PV as the control unit. 

\subsection{Formulation of Markov Games in Active Voltage Control}
The collaborative control process of PV inverters can be modeled as a Dec-POMDP\cite{Dec-POMDP} for $N$ agents. A Dec-POMDP is usually defined by a tuple $(\mathcal{N},\mathcal{S}, \{ \mathcal{A}_i \} _{i \in \mathcal{N}}, \mathcal{T}, r, \{ \mathcal{O}_i \} _{i \in \mathcal{N}}, \Omega, \gamma)$, where $\mathcal{N} = \{ 1, ... , n\}$ denotes the set of $n$ agents, $\mathcal{S}$ denotes the state space observed by all agents, $\mathcal{A}_i$ denotes the action space of agent $i$. Let $\mathcal{A} = \times_{i \in \mathcal{N}} \mathcal{A}_i$, then $\mathcal{T} : \mathcal{S} \times \mathcal{A} \times \mathcal{S} \rightarrow [0,1]$ denotes the transition probability from any state $s \in \mathcal{S}$ to any state $s' \in \mathcal{S}$ after taking a joint action $a \in \mathcal{A}$; \ $r : \mathcal{S} \times \mathcal{A} \rightarrow \mathbb{R}$ is a global reward function that determines the immediate reward received by whole agents for a transition from $(s,a)$ to $s'$; $\mathcal{O} = \times_{i \in \mathcal{N}} \mathcal{O}_i$ denotes the joint observation set, where $\mathcal{O}_i$ is each agent's observation; $\Omega : \mathcal{S} \times \mathcal{A} \times \mathcal{O} \rightarrow [0,1]$ denotes the perturbation of the observers for agents' joint observations over the states after decisions; $\gamma \in [0,1)$ is the discount factor. We can formulate the policy of the i-th agent's policy as $\pi^i$,  and the objective of Dec-POMDP is finding an optimal joint policy $\pi = \times_{i \in \mathcal{N}} \pi^i$ to maximize expected long-term reward $\mathbb{E}_{\pi} [ \sum_{t=0}^\infty \gamma^t r_t]$. Considering the active voltage control problem, we describe specific elements in the Dec-POMDP in detail as follows, simulated to \cite{MAPDN}:

\begin{itemize}[leftmargin=*,topsep=0pt,parsep=0pt]
\item \textbf{Agent.} As shown in Figure \ref{fig:illustration}, each PV is an agent that injects the reactive power generated by it's PV inverter into the distribution network so as to maintain the voltage of all buses within the safe range.
% \item \textbf{Observation.} Let $o_j = (p^L_j, q^L_j, v_j, \theta_j, flag^{PV}_j)$ represent the feature of the node $j$. $p^L_j \in (0, \infty)$ and $q^L_j \in (0,\infty)$ are active and reactive power of the load connected to the node $j$ respectively; $v_j \in (0,\infty)$ and $\theta_j \in [-\pi, \pi]$ denote the voltage magnitude and voltage phase of the node $j$ respectively; $flag^{PV}_j \in \{ 0, 1\}$ indicates whether the node $j$ is installed with a PV. Thus, the observation of agent $i$ is defined as $\mathcal{O}_i = (\mathbf{f}, p^{PV}, q^{PV})$. $\mathbf{f} = \{ o_j $ $|$ the PV $i$ and the node $j$ are in common zone $\}$ is the set of information about the nodes in the same zone as PV $i$; $p^{PV} \in (0, \infty)$ is the active power generated by PV $i$; and $q^{PV} \in (-\infty, \infty)$ is the reactive power generated by PV inverter $i$ at the preceding step.
\item \textbf{Observation.} Let $o_j = (p^L_j, q^L_j, v_j, \theta_j, flag^{PV}_j)$ represents the node-based feature of node $j$. $p^L_j \in (0, \infty)$ and $q^L_j \in (0,\infty)$ are active and reactive power of the load connected to node $j$ respectively; $v_j \in (0,\infty)$ and $\theta_j \in [-\pi, \pi]$ denote the voltage magnitude and voltage phase of node $j$ respectively; $flag^{PV}_j \in \{ 0, 1\}$ indicates whether the node $j$ is installed with a PV. Moreover, additional physical quantities $(p^{PV}, q^{PV})$ are appended to the node-based feature for those nodes installed with PV. $p^{PV} \in (0, \infty)$ denotes the active power generated by PV $i$ and $q^{PV} \in (-\infty, \infty)$ is the reactive power generated by the PV inverter. Nodes in the grid are partitioned into several zones and the observation of agent $i$ (denoted as $\mathcal{O}_i$) is obtained by concatenating node-based features in the zone in which agent $i$ is located. It is worth noting that agents in the same zones have similar observations.
\item \textbf{Action.} Each agent $i \in \mathcal{N}$ has a continuous action set $\mathcal{A}_i = \{ a_i : -c \leq a_i \leq c, c>0\}$ that denotes the ratio of maximum reactive power it can generate. 
% Then, the reactive power generated by the agent (PV) $i$ is calculated as follows: $q^{PV}_i = a_i \sqrt{(s^{max}_i)^2 - (p^{PV}_i)^2}$. $s^{max}_i$ is the maximum apparent power, depending on physical capacity of the PV inverter. If $a_i > 0$, it means penetrating reactive powers to the distribution network. Conversely, if $a_i < 0$, it means absorbing reactive powers from the distribution network. Typically, the value of $c$ is chosen in accordance with the load capacity of distribution network for safety. 
And the joint action set is defined as $\mathcal{A} = \times _{i \in \mathcal{N}} \mathcal{A}_i$.
\item \textbf{Reward Function.} The reward function is defined as follows: 
\begin{equation}
\label{equ:reward_function}
r=-\frac{1}{|V|} \sum_{i \in V} l_{v}\left(v_{i}\right)-\alpha \cdot l_{q}\left(\mathbf{q}^{PV}\right),
\end{equation}
where $l_v(\cdot)$ is a voltage barrier function and $l_{q} = \frac{1}{|\mathcal{N}|}\left\|\mathbf{q}^{P V}\right\|_{1}$ is the reactive power generation loss. The objective is to learn a optimal strategy to control the voltage within the safety range around $V_{ref}$ (i.e. $[0.95v_{ref}, 1.05v_{ref}]$) while minimizing the power loss of the whole distribution network. The hyper-parameter $\alpha \in (0,1)$ is set in advance by simulation environment, which plays the role of balancing the two losses. In practice, we use the reactive power generation loss instead of power loss of the whole distribution network, because obtaining overall power loss of the grid in real time is difficult. The voltage barrier function $l_v(\cdot)$ penalizes the voltage rise deviation and the voltage drop deviation. As with \cite{MAPDN}, there voltage barrier functions are considered to form reward functions: L1-shape, L2-shape and bowl-shape (see Appendix \ref{sec:function}).
\end{itemize}

Additionally, the topology of the grid will be exploited as a priori knowledge in our framework.
Formally, the connectivity between nodes in a zone is represented by a adjacency matrix $D^i$.
% Formally, the adjacency matrix $D^i$ indicates how nodes are connected in the zone where agent $i$ is located. 
% If node $x$ and node $y$ are adjacent, $D^i_{xy}$ equals $1$; If they are not, $D^i_{xy}$ equals $0$.
Furthermore, we divide the raw observation of agent $i$ into node-based features based on the prior knowledge:
\begin{equation}
\label{equ:obs}
\mathcal{O}_i = \{ o_{i,1}, o_{i,2}, \cdots, o_{i,m_i} \}.
\end{equation}
Here, $i \in \{ 1,2,\cdots,n \}$ is the index of agents, $m_i$ denotes the number of nodes in the zone.
Let $p_i \in \{ 1,2,\cdots,m_i\}$ denotes the index of node installed with PV $i$, which makes agent $i$ aware of its own location in the zone.

\section{Method}
\label{sec:method}
In order to apply MARL algorithms to the active voltage control problem while considering the characteristics of grid, we present a novel transformer-based multi-agent actor critic (T-MAAC) framework.
Specifically, we propose a policy network and a critic network based on transformer as well as auxiliary-task based training process, which is compatible with mainstream multi-agent actor-critic algorithms such as MADDPG\cite{MADDPG} and MATD3\cite{MATD3}.
In this section, we describe the structure of the policy network and the critic network in Section \ref{sec:policy} and Section \ref{sec:critic}, respectively.
The auxiliary task to stabilize the training process is discussed in Section \ref{sec:auxiliary}.

% \subsection{Motivation}
% \label{sec:motivation}
% Among most of MARL algorithms in fully collaborative environment, all observations are padded to the same dimension and the parameters of policy networks of different agents are shared in order to increase the scalability(i.e. \cite{Cooperative,MADDPG,MATD3,MAPPO,COMA}).
% However, the composition of observations varies greatly from agent to agent in active voltage control problem, making it difficult for the shared policy network to adapt to various raw padded observations. Inspired by \cite{transformer}, we further design a transformer-based policy network suitable for active voltage control problem to handle various observation sizes in Section \ref{sec:policy}.

%% TODO: How to intorduce the motivation of transofmer-based critic netwrok
% Multi-agent credit assignment technique is also significant during training process in cooperative scenarios. 

\subsection{Transformer-based Policy Network}
\label{sec:policy}
\begin{figure}%[h]
  \centering
  \includegraphics[width=\linewidth]{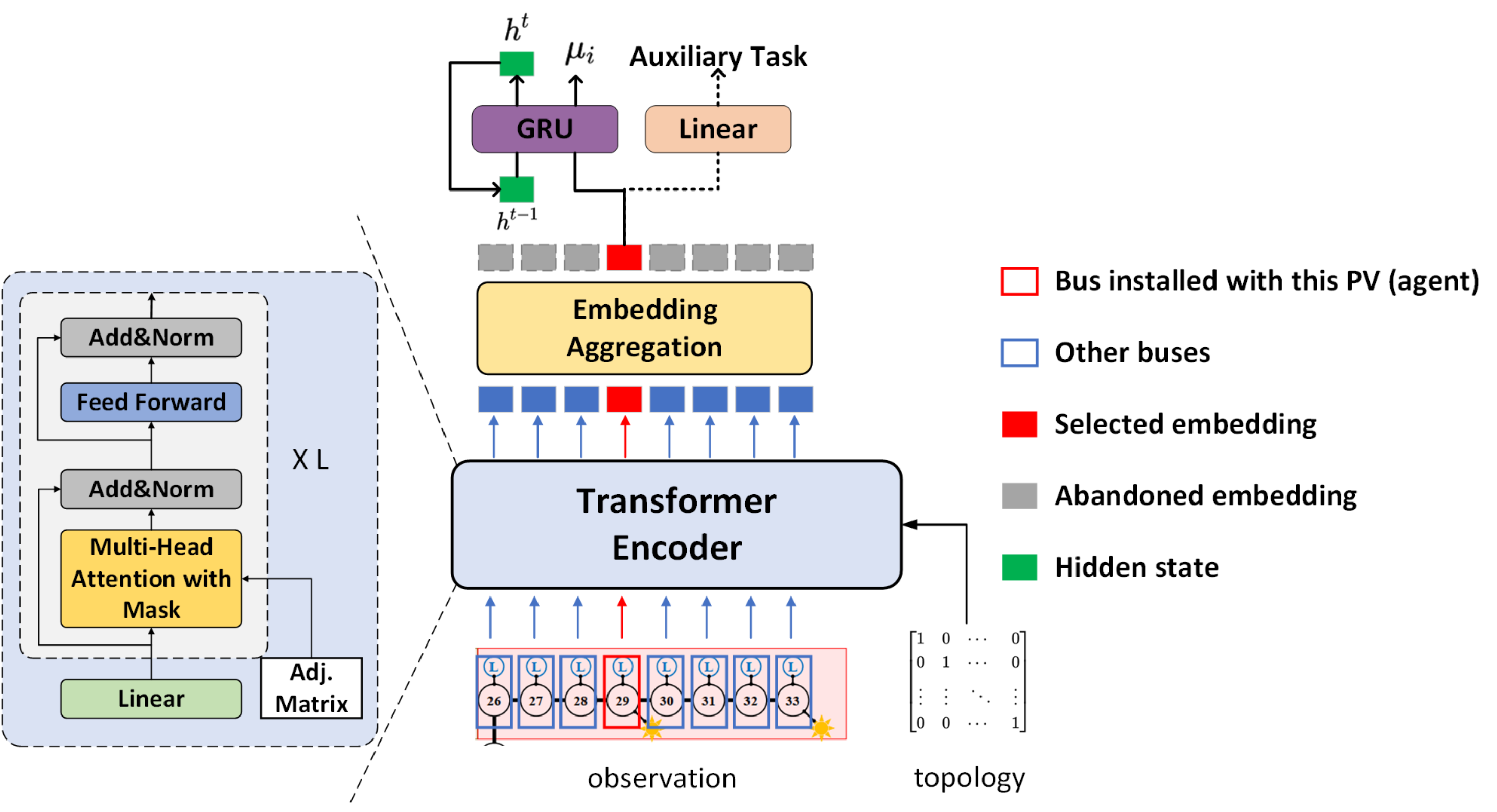}
  \caption{The architecture of the policy network based on transformer. The transformer encoder and embedding aggregation module are used to obtain the embedding of raw observation. Then, the embedding is mapped to action via the GRU head. Additionally, the auxiliary-task head predicts the voltage out of control ratio \textit{(VR)} of the raw observation, which stabilizes the training process via extra auxiliary loss. Details can be found in Section \ref{sec:policy}.}
  \label{fig:policy_network}
\end{figure}
% Among most of MARL algorithms, all observations are padded to the same dimension and various agents share the parameters of policy networks(\cite{Cooperative,MADDPG,MATD3,MAPPO,COMA}).
% Conventionally, multi-layer perceptions are used to extract features that are relevant to the particular task.
% However, this simple architecture can not deal with nature in grid as following: 
% \begin{enumerate}[leftmargin=*]
% \item \textbf{Inconsistent number of nodes observed by various agents.}
% \item \textbf{Inconsistent topology of nodes in each observation.}
% \item \textbf{Inconsistent importance of nodes in a zone.}
% \end{enumerate}
To extract more relevant representations for the active voltage control task, we develop a transformer-based policy network as shown in Figure \ref{fig:policy_network} to handle various types of observations. We present a mathematical formulation of our transformer-based model in this section.

\subsubsection{Projection layer}
First of all, we transform the raw observation $\mathcal{O}$ into node-based embeddings via a projection layer.

If the observation of agent $i$ at time step $t$ (denoted as $\mathcal{O}^t_i$) is made up of $m$ node-based features, then all of them are embedded via a projection layer $P$ as follows:
\begin{equation}
\label{equ:embedding}
input_i^t = \{ P(o^t_{i,1}), P(o^t_{i,2}), \cdots, P(o^t_{i,m_i})\}.
\end{equation}
In Eq.~\eqref{equ:embedding}, $i \in \{ 1, \cdots, n\}$ is the index of the agent; $j \in \{ 1,  \cdots, m_i\}$ is the index of nodes; $m$ denotes the number of nodes in the zone.

In the vanilla transformer\cite{transformer}, Vaswani adds "positional encodings" to the input embeddings to inject position information of the tokens in a sequence. However, a distribution network has a radial topology instead of sequential structure. So instead of positional encodings, we inject position information via an adjacency matrix $\mathbf{D}^i$ used in the attention mechanism that will be elaborated in the next section.

\subsubsection{Transformer Encoder and Embedding Aggregation Module}
Next, The transformer encoder and embedding aggregation module are designed to extract more robust representation from the $input^t_i$ above as shown in Figure \ref{fig:policy_network}.
% As shown in Figure \ref{fig:policy_network}, multiple identical transformer layers\cite{transformer} are stacked together to form the transformer encoder. Each layer is composed of four sub-layers: a multi-head self-attention with mask sub-layer, a fully connected feed-forward sub-layer, and two layer-normalization sub-layers.

The vanilla self-attention mechanism proposed in \cite{transformer} is computed as follows:
\begin{equation}
\label{equ:self-attention}
    \operatorname{Attention} \left(\mathbf{Q},\mathbf{K},\mathbf{V} \right) = \operatorname{softmax}\left(\frac{\mathbf{Q K}^{T}}{\sqrt{d_{k}}}\right) \mathbf{V}.
\end{equation}
Eq.~\eqref{equ:self-attention} can be extended to self-attention with mask mechanism:
\begin{equation}
\label{equ:self-attention-mask}
    \operatorname{MaskAttention} (\mathbf{Q},\mathbf{K},\mathbf{V},\mathbf{mask}) = \operatorname{softmax}\left(\frac{\mathbf{Q K}^{T}}{\sqrt{d_{k}}} \cdot \mathbf{mask}\right) \mathbf{V}.
\end{equation}
Three matrices $\mathbf{K}, \mathbf{Q}, \mathbf{V}$ represent a set of keys, queries and values respectively; $d_k$ is a scaling factor equal to dimension of queries and keys; $\mathbf{mask}$ is a binary matrix with the same shape as $\mathbf{Q K}^T$ and $mask_{x,y} \in \{ 0, 1\}$ indicates whether perform an attention operation between position $x$ and position $y$.
%In our method, the adjacency matrix $\mathbf{D}^i$, which indicates how nodes are connected, is chosen as the $\mathbf{mask}$.

Our transformer encoder utilizes the self-attention with mask mechanism to establish the correlation between nodes in the zone. We formulate our transformer encoder as follows:
\begin{align}
E^{0} &= input^t_i, \\
\label{equ:att1}
Q^{(l)}, K^{(l)}, V^{(l)} &=W_{Q}^{(l)} E^{(l-1)}, W_{K}^{(l)} E^{(l-1)}, W_{V}^{(l)} E^{(l-1)}, \\
\label{equ:att2}
\bar{Y}^{(l)} &= \operatorname{MaskAttention} \left(  Q^{(l)}, K^{(l)}, V^{(l)}, D^i \right), \\
\label{equ:att3}
E^{(l)} &=\operatorname {LayerNorm}\left(E^{(l-1)}+\operatorname{Linear}\left(\bar{Y}^{(l)}\right)\right),
\end{align}
where $W_Q^{(l)}, W_K^{(l)}, W_V^{(l)}$ represent the learnable parameters to compute $\mathbf{Q}, \mathbf{K}, \mathbf{V}$ and $l \in \{ 1, 2, \cdots, L \}$ is the index of transformer layers.
The adjacency matrix $D^i$ indicates how nodes are connected.
For example, if node $x$ and node $y$ are adjacent, $D^i_{xy}$ equals $1$; if they are not, $D^i_{xy}$ equals $0$.

Then, the embedding aggregation module aggregates $E^{(l)}$ into global information $\bar{E}_i$ from the view of agent $i$. In practice, we select an additional transformer layer without mask as our embedding aggregation module:
\begin{align}
Q^{EA}, K^{EA}, V^{EA} &=W_{Q}^{EA} E^{(l)}, W_{K}^{EA} E^{(l)}, W_{V}^{EA} E^{(l)}, \\
\label{equ:att4}
\bar{Y}^{EA} &= \operatorname{Attention} \left(  Q^{EA}, K^{EA}, V^{EA} \right), \\
\label{equ:att5}
E^{EA} &=\operatorname {LayerNorm}\left(E^{(l)}+\operatorname{Linear}\left(\bar{Y}^{EA}\right)\right), \\
\label{equ:att6}
\bar{E}_i &= \operatorname{SelectEmbedding} \left( E^{EA}, p_i \right).
\end{align}
$\operatorname{SelectEmbedding}$ is an operation to select the embedding whose index is $p_i$ (The index of node installed with PV $i$). 
Intuitively, $\bar{E}_i$ denotes the representation extracted from the agent $i$ perspective in a higher semantic space.

\subsubsection{GRU head and Auxiliary-task head}
In the last part of the policy network, a GRU\cite{GRU} layer is applied to project the representation $\bar{E}_i$ to the control action $\mu_i$. $h^t$ in Figure \ref{fig:policy_network} denotes the temporal hidden state at the time step $t$.

Meanwhile, we introduce the auxiliary-task head to predict the voltage
out of control ratio \textit{(VR)} in $\mathcal{O}_i$. \textit{VR} indicates the ratio of voltage outside the safety range (i.e. 0.95-1.05 p.u.) in a zone, which is a critical metric that corresponds to the optimization objectives for active voltage control task.
Thus, the auxiliary-task head helps the upstream transformer encoder module implicitly recover this essential information from raw observation. 
Also, extra auxiliary loss helps stabilize the learning process, which is introduced in Section \ref{sec:auxiliary} in detail.

\subsection{Transformer-based Critic Network}
\label{sec:critic}
\begin{figure}%[h]
  \centering
  \includegraphics[width=0.8\linewidth]{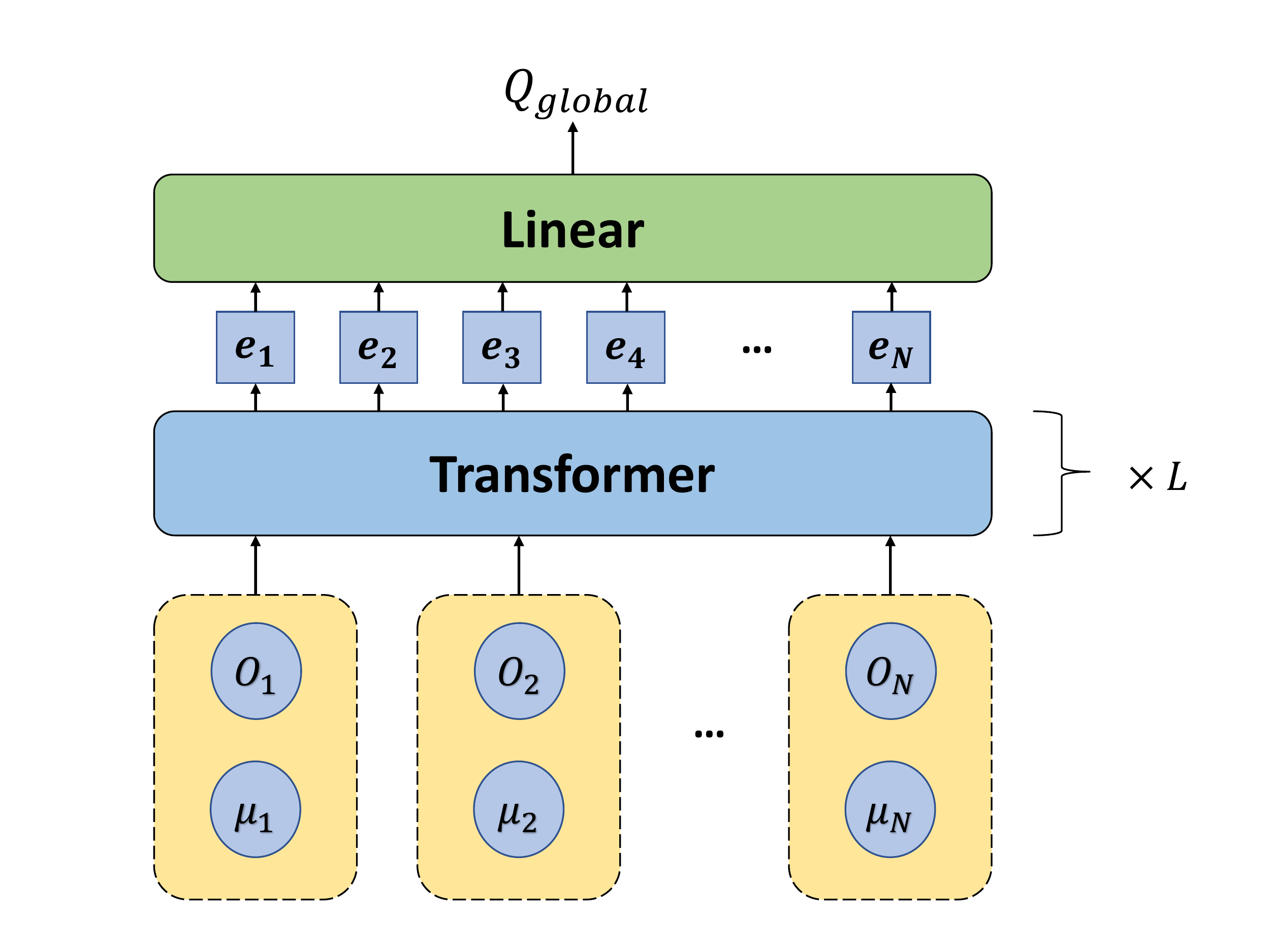}
  \vspace{-4mm}
  \caption{The architecture of the global critic network. Our framework replace the widely used MLP-based critic network with the transformer-based critic network.}
  \label{fig:critic_network}
\end{figure}
For the active voltage control task, the correlations between agents are related to their position in the grid.
In a radial distribution network, the voltage of each node is influenced by all other nodes, but the impact decreases as the distance increases. Therefore, agents need to be aware of the relationship between nodes in the grid to make cooperative control decisions.
For example, if two agents are topologically close to each other, they must consider more carefully when making decisions.
Additionally, the voltage of most nodes can be controlled within the safety range (i.e. from $0.95 p.u.$ to $1.05 p.u.$), while only a few parts have the risk of voltage exceeding the safety value. Thus, agents may pay more attention to those zones in danger to control the voltage of all nodes in the distribution network.

The conventional centralized critic network constructed with pure MLPs\cite{MAPDN} suffers from credit assignment issues, especially in a large-scale cooperative environment.
The large number of agents and their complex relationships complicate policy learning\cite{MADDPG_GNN}.
To make the learning process more robust, we model the correlation between agents via the self-attention mechanism and design a transformer-based critic network. The architecture of critic network is shown in Figure \ref{fig:critic_network}.

We describe the calculation of the global Q-function parameterized by the transformer-based critic network. 
First of all, let $\mathcal{O}_i$ denotes the raw observation of agent $i$ and $\mu_i$ denotes the corresponding action it performed. Then, $N$ tuples $(\mathcal{O}_i, \mu_i)$ are transformed into $N$ embeddings through several vanilla transformer layers\cite{transformer}:
\begin{equation}
\{ e_1,\cdots, e_N \} = \operatorname{Transformer} \left(\{  (\mathcal{O}_1, \mu_1), \cdots, (\mathcal{O}_N, \mu_N) \} \right).
\end{equation}
Next, we project the embeddings to the output space of the centralized action-value function $Q^{\pi}_{global}$ through a linear function:
\begin{equation}
Q^{\pi}_{global}(\mathcal{O}_1, \mu_1, \mathcal{O}_2, \mu_2, \cdots, \mathcal{O}_N, \mu_N) = \operatorname{Linear}(e_1, e_2, \cdots, e_N).
\end{equation}
In addition, our proposed transformer-based critic network reduces the number of parameters that benefits from compact semantic representations calculated by transformer architecture.

\subsection{Auxiliary-task Training Process}
\label{sec:auxiliary}
% Our proposed transformer-based multi-agent actor-critic (T-MAAC) method is a generic framework, which is compatible with various multi-agent actor-critic algorithms such as MADDPG\cite{MADDPG}, MATD3\cite{MATD3} etc.
In this section, we select MADDPG\cite{MADDPG} as the base algorithm to describe the entire auxiliary-task training process of T-MAAC.
Suppose there are a total of $N$ agents with continuous policies $ \boldsymbol{\mu}_i (\cdot \ ; \theta_i)$ parameterized by $\mathbf{\theta} = \{ \theta_1, \cdots, \theta_N\}$.
Let $\mathbf{s} = (o_1, \cdots, o_N)$ denotes the observations of each agents, then we formulate the gradient of the expected return for agent $i$, $J(\boldsymbol{\mu}_i) = \mathbb{E}[R]$ as:
\begin{equation}
\label{equ:actor_loss}
\resizebox{.9\hsize}{!}{
$\nabla_{\theta_{i}} J\left(\boldsymbol{\mu}_{i}\right)=\mathbb{E}_{\mathbf{s}, a \sim \mathcal{D}}\left[\left.\nabla_{\theta_{i}} \boldsymbol{\mu}_{i}\left(a_{i} \mid o_{i}\right) \nabla_{a_{i}} Q_{global}^{\boldsymbol{\mu}}\left(\mathbf{s}, a_{1}, \ldots, a_{N}\right)\right|_{a_{i}=\boldsymbol{\mu}_{i}\left(o_{i}\right)}\right]$.
}
\end{equation}
Here, $\mathcal{D}$ is the experience replay buffer recording transitions of all agents.
And the centralized global action-value functions $Q^{\boldsymbol{\mu}}_{global}$ is updated as:
\begin{align}
\quad y &= r+\left.\gamma Q_{global}^{\boldsymbol{\mu}^{\prime}}\left(\mathbf{s}^{\prime}, a_{1}^{\prime}, \ldots, a_{N}^{\prime}\right)\right|_{a_{j}^{\prime}=\boldsymbol{\mu}_{j}^{\prime}\left(o_{j}\right)}, \\
\mathcal{L}\left(\theta_{i}\right) &= \mathbb{E}_{\mathbf{s}, a, r, \mathbf{s}^{\prime}}\left[\left(Q_{global}^{\boldsymbol{\mu}}\left(\mathbf{s}, a_{1}, \ldots, a_{N}\right)-y\right)^{2}\right].
\label{equ:critic_loss}
\end{align}
In addition, we adopt an additional self-supervised loss to stabilize the learning process.
$\mathbf{v}_i (\cdot \ ; \theta_i)$ is the output of auxiliary-task head in the policy network, which predicts the voltage out of control ratio \textit{(VR)} in the zone.
And it's ground truth $label_i$ is calculated from the raw observation $o_i$.
The auxiliary loss is as follows:
\begin{equation}
label_i = \sum_{j=1}^{m_i} [ \mathbb{I}(Voltage_j < 0.95) + \mathbb{I}(Voltage_j > 1.05) ], \end{equation}
\begin{equation}
\label{equ:aux_loss}
\mathcal{L}_{aux}\left(\theta_{i}\right) = \mathbb{E}_{\mathbf{s}, a, r, \mathbf{s}^{\prime}}\left[ (\mathbf{v}_i \left(o_i \ ; \theta_i)  - label_i \right)^2 \right],
\end{equation}
where $\mathbb{I}$ is the indicator function and $m_i$ is the number of nodes in the $o_i$.
In order to improve the sample efficiency and scalability in MARL algorithms, the parameters of policy networks are shared among agents in T-MAAC. 
Details of transformer-based MADDPG (T-MADDPG) are given in Algorithm \ref{alg:alg1}.

\begin{algorithm}
\caption{Transformer-based MADDPG (T-MADDPG)}
\label{alg:alg1}
\BlankLine
\For{episode $\leftarrow 1$ \KwTo $M$}{
    Initialize a random process $\mathcal{N}$ for action exploration and Replay Buffer $\mathcal{D}$\;
    \For{t $\leftarrow 1$ \KwTo max-episode-length}{
        for each agent $i$, select a action $a_i = \mu_{\theta_i}(o_i) + \mathcal{N}_i$\;
        Execute actions $a=(a_1, \cdots, a_N)$ from state $s$\;
        Get reward $r$ by going to new state $s'$\;
        Store $(s,a,r,s')$ in replay buffer $\mathcal{D}$ and $s \leftarrow s'$\;
        \For{agent $i \leftarrow 1$ \KwTo N}{
            Sample a random mini-batch of $\mathcal{S}$ samples $(s^j, a^j, r^j, s'^j)$ from $\mathcal{D}$ \;
            Update critic by minimizing the loss in Eq. (\ref{equ:critic_loss}) \;
            Update actor by minimizing the auxiliary loss in Eq. (\ref{equ:aux_loss}) \;
            Update actor by gradient ascent in Eq. (\ref{equ:actor_loss}) \;
        }
        Update target network parameters for each agent $i$: $\theta'_i \leftarrow \tau\theta_i + (1-\tau)\theta'_i$\;
    }
}
\end{algorithm}

\begin{figure*}%[htbp]
    \centering
    \subfigure[CR-L1-141]{
        \includegraphics[scale=0.12]{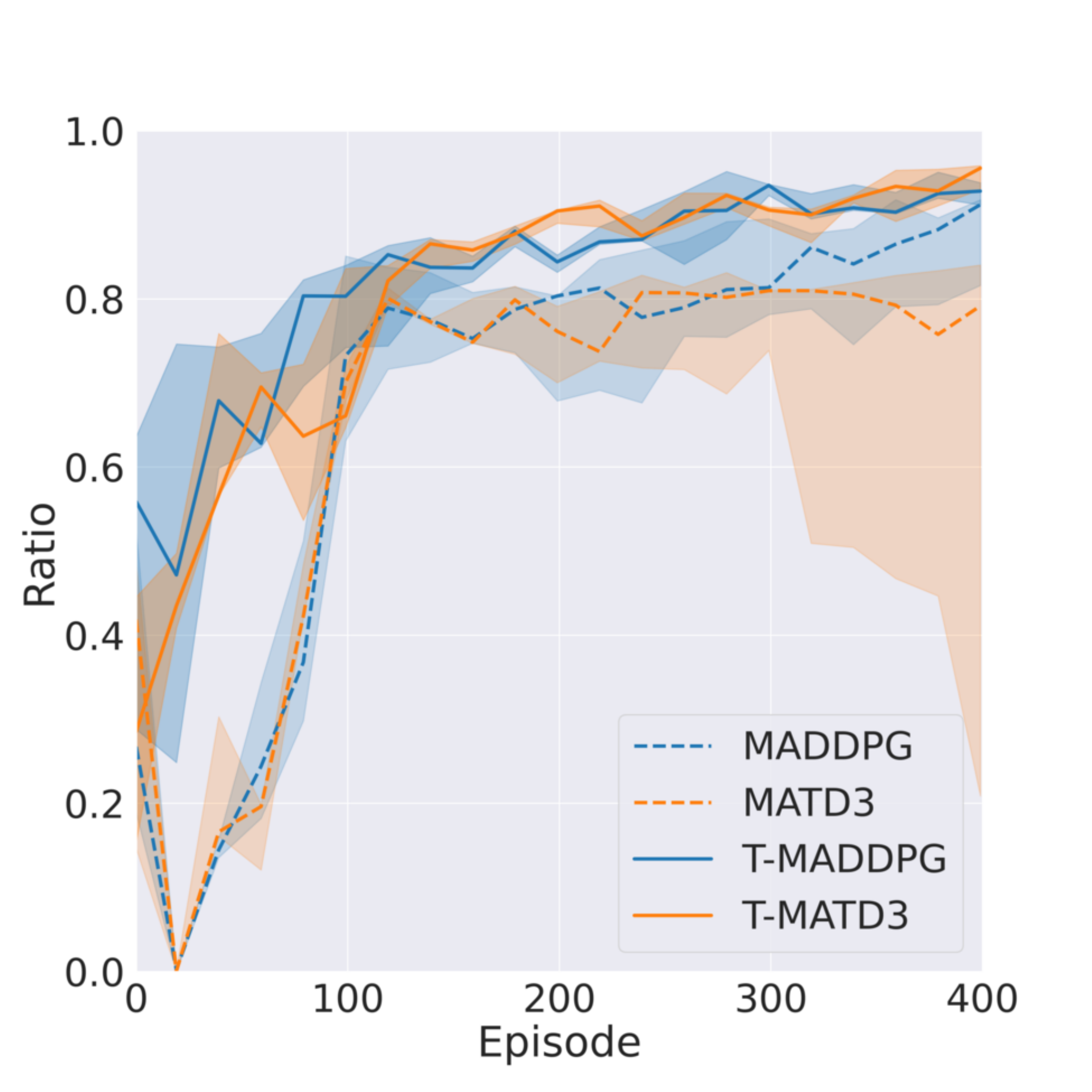}
    }
    \subfigure[CR-L2-141]{
        \includegraphics[scale=0.12]{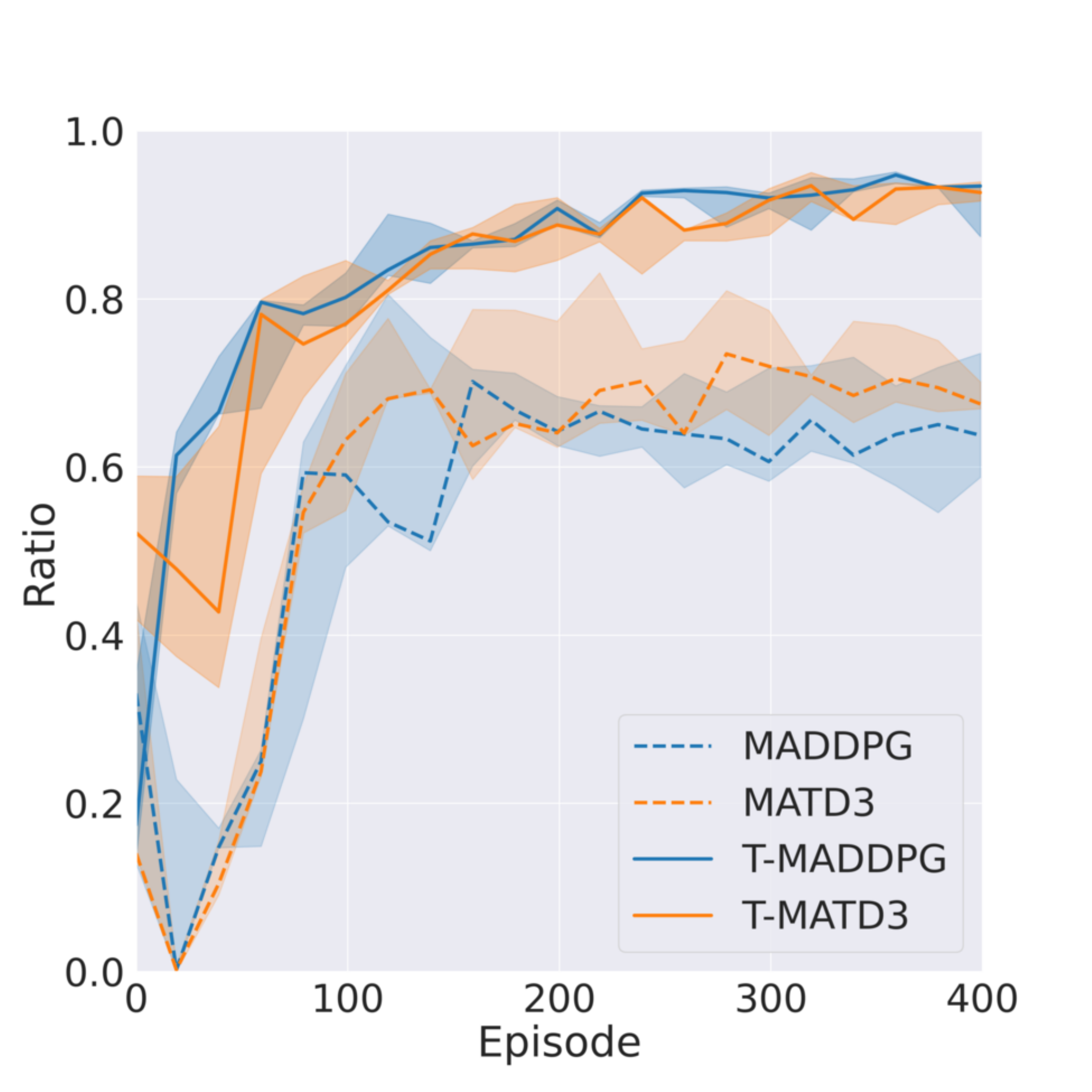}
    }
    \subfigure[CR-BL-141]{
        \includegraphics[scale=0.12]{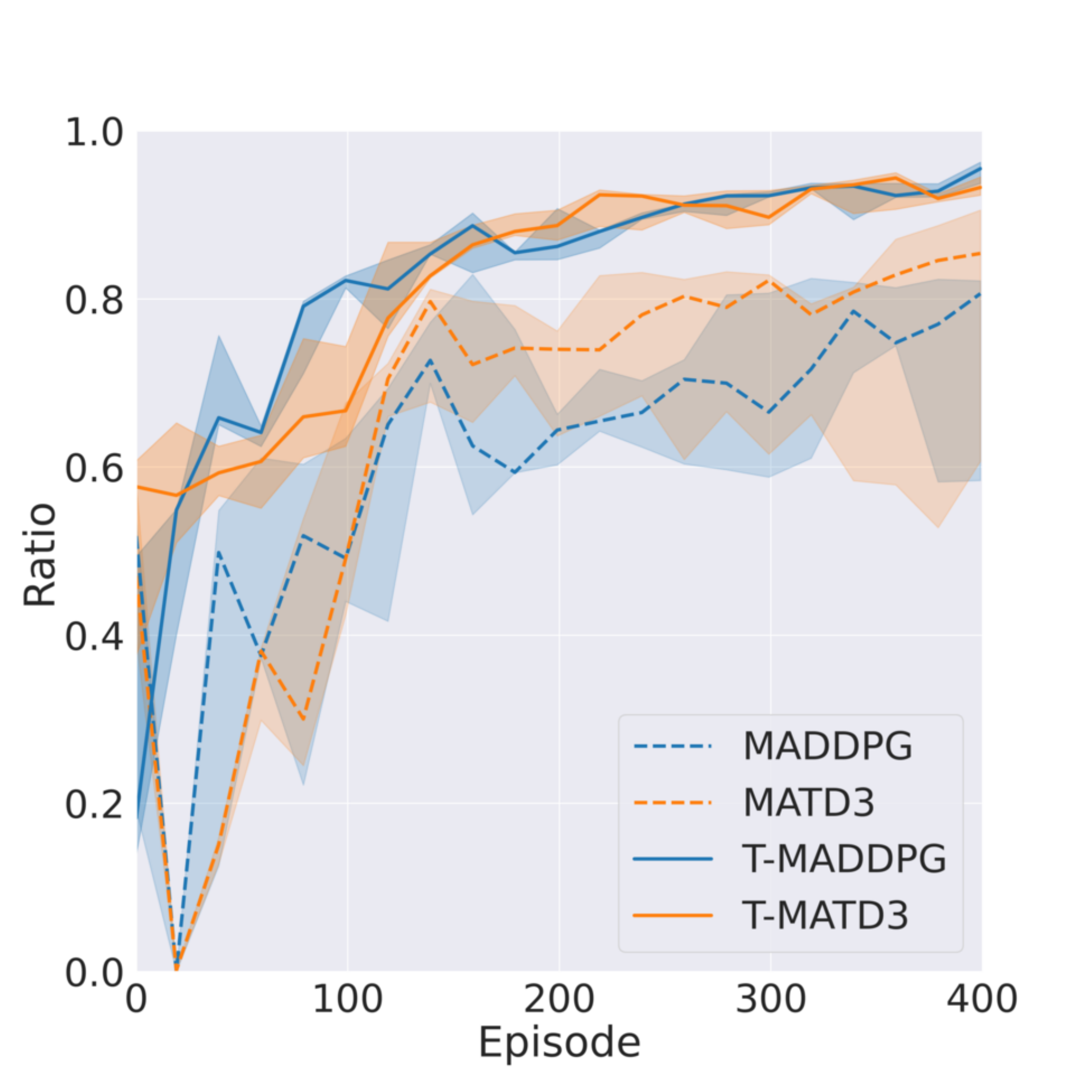}
    }
    \subfigure[CR-L1-322]{
        \includegraphics[scale=0.12]{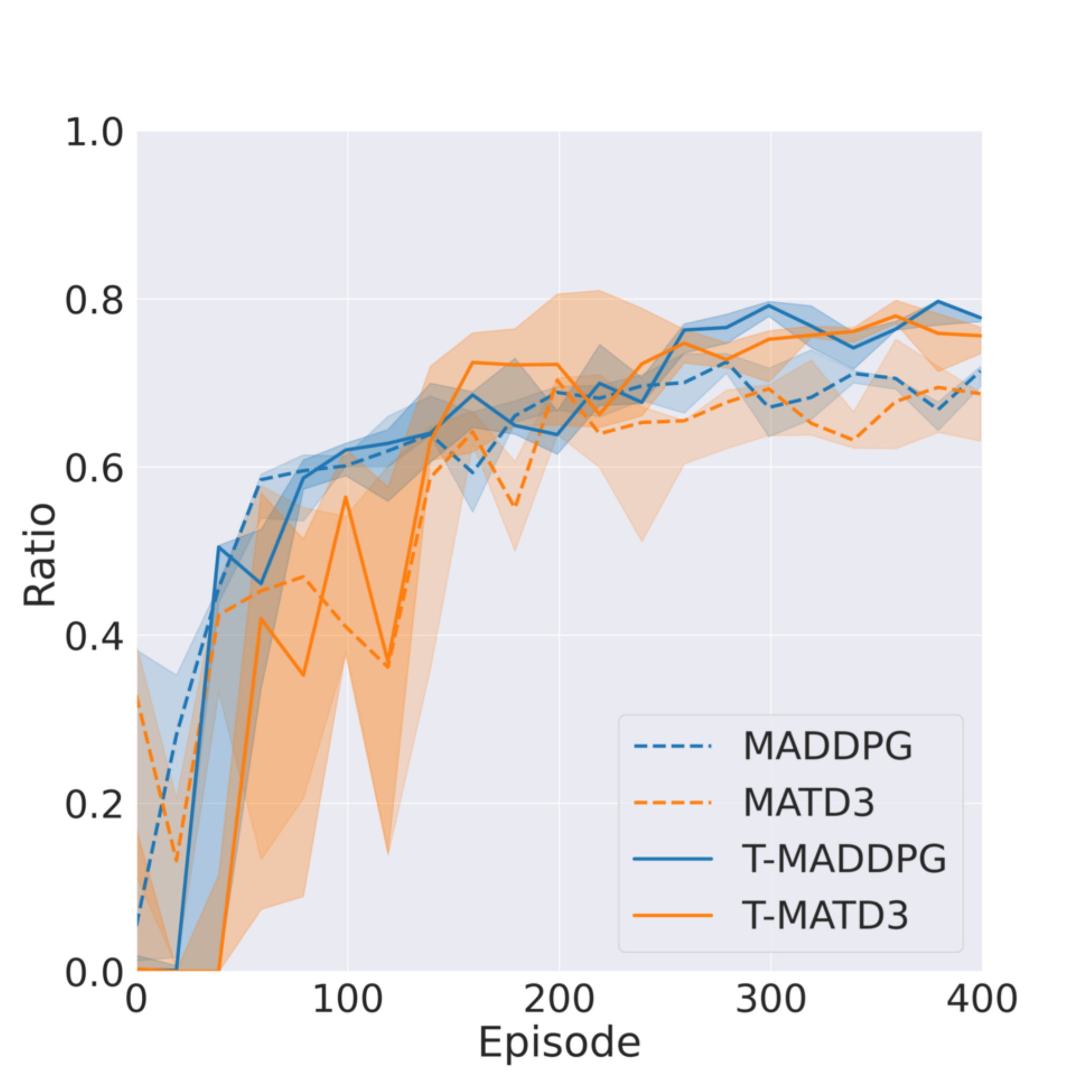}
    }
    \subfigure[CR-L2-322]{
        \includegraphics[scale=0.12]{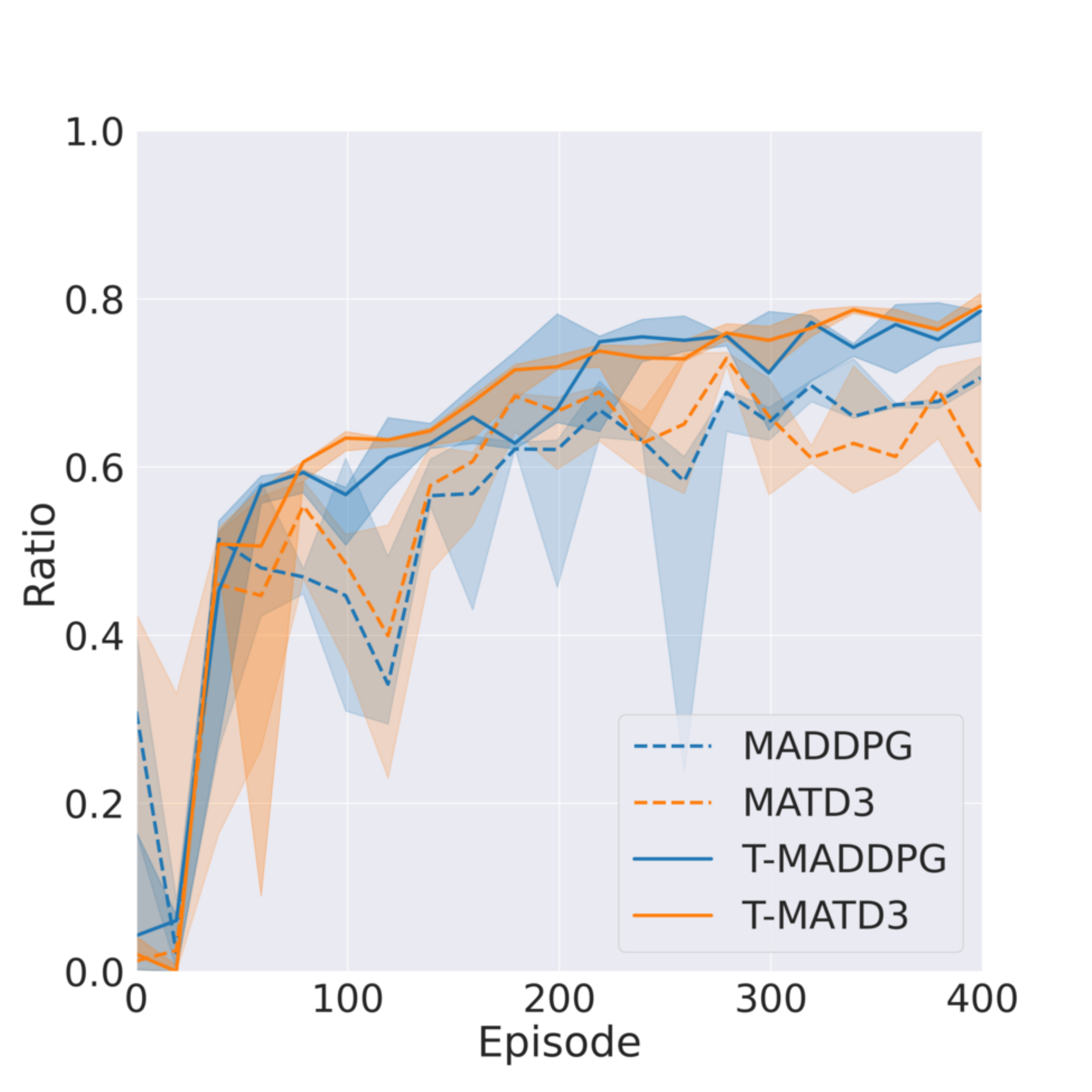}
    }
    \subfigure[CR-BL-322]{
        \includegraphics[scale=0.12]{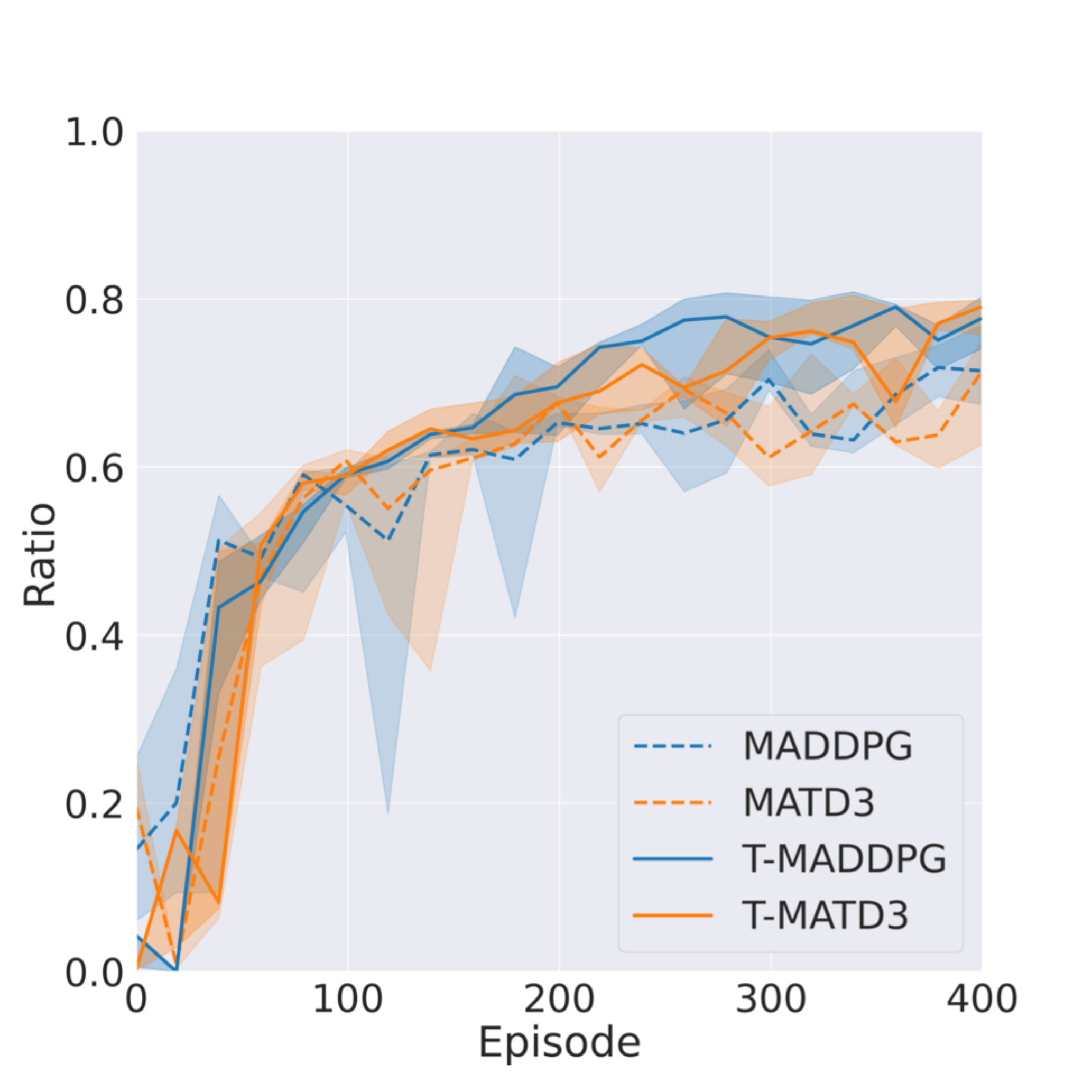}
    }\vspace{-4mm}
    \\
    \subfigure[QL-L1-141]{
        \includegraphics[scale=0.12]{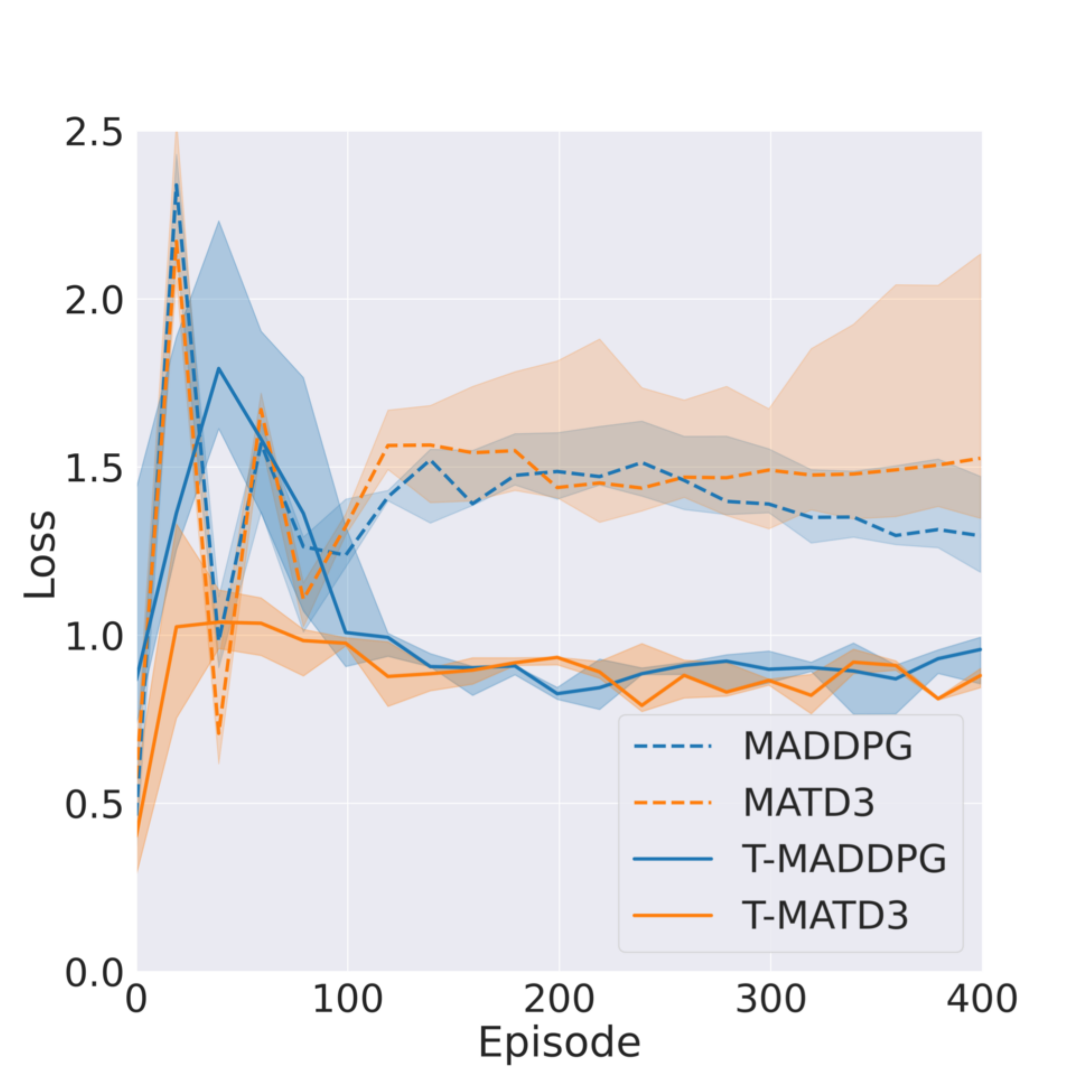}
    }
    \subfigure[QL-L2-141]{
        \includegraphics[scale=0.12]{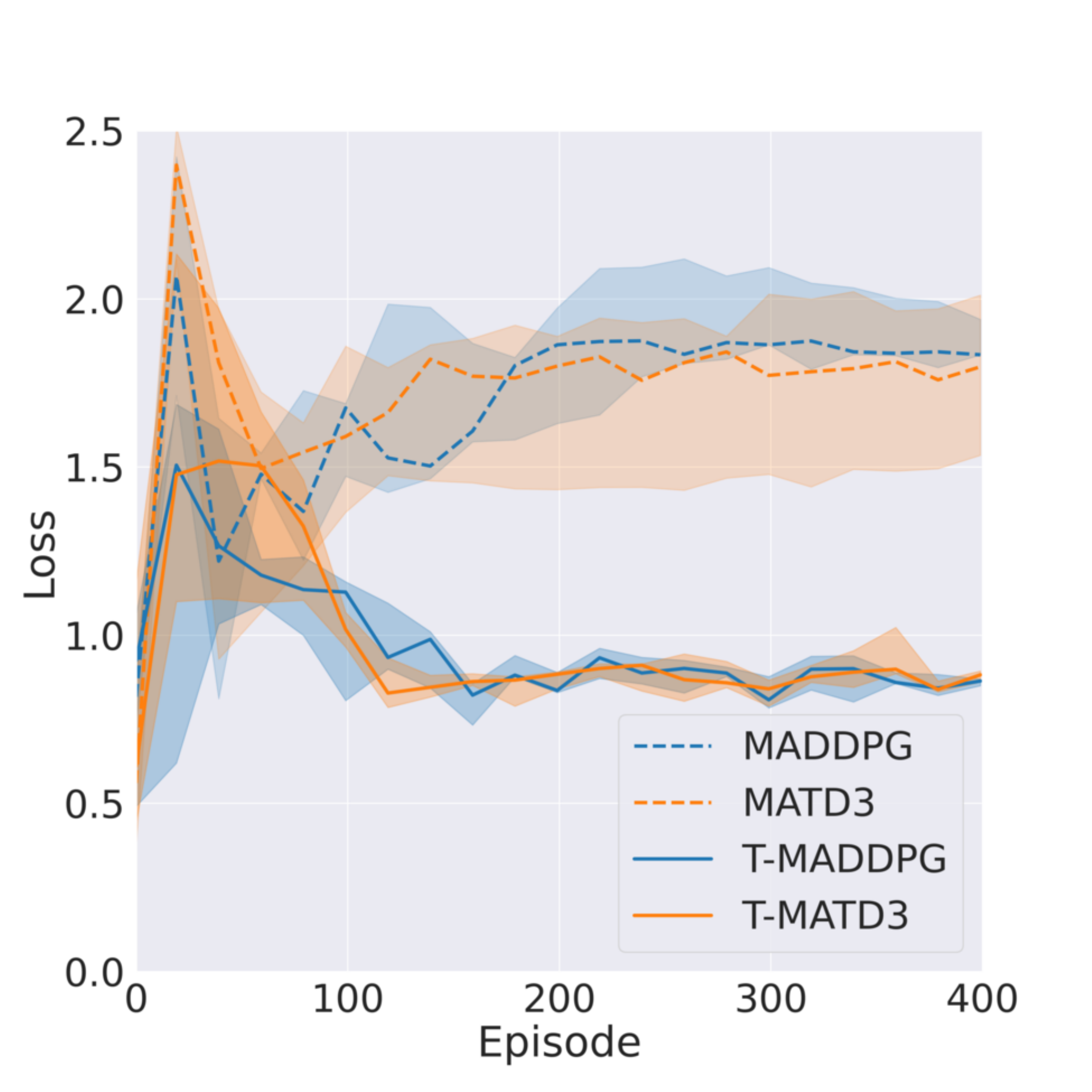}
    }
    \subfigure[QL-BL-141]{
        \includegraphics[scale=0.12]{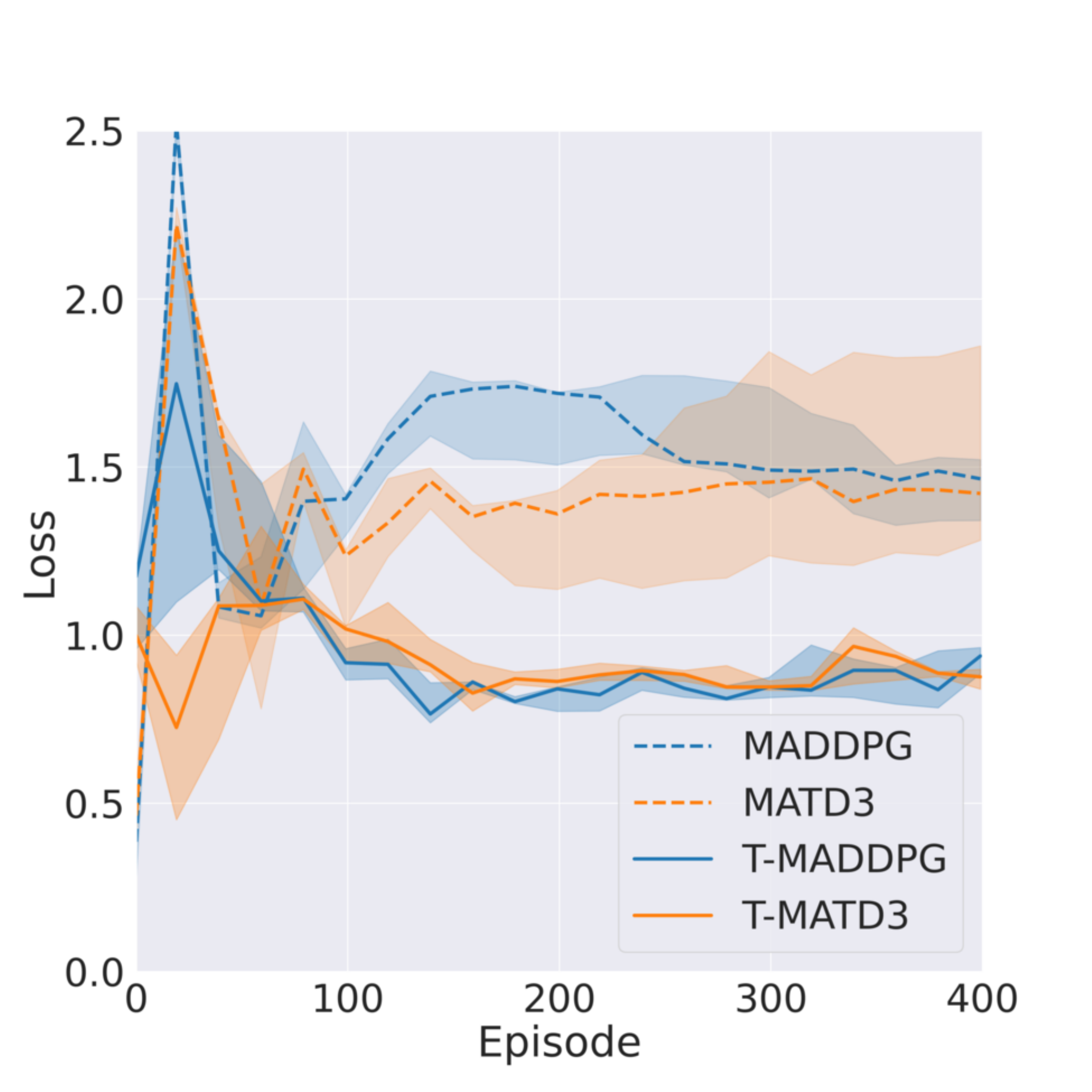}
    }
    \subfigure[QL-L1-322]{
        \includegraphics[scale=0.12]{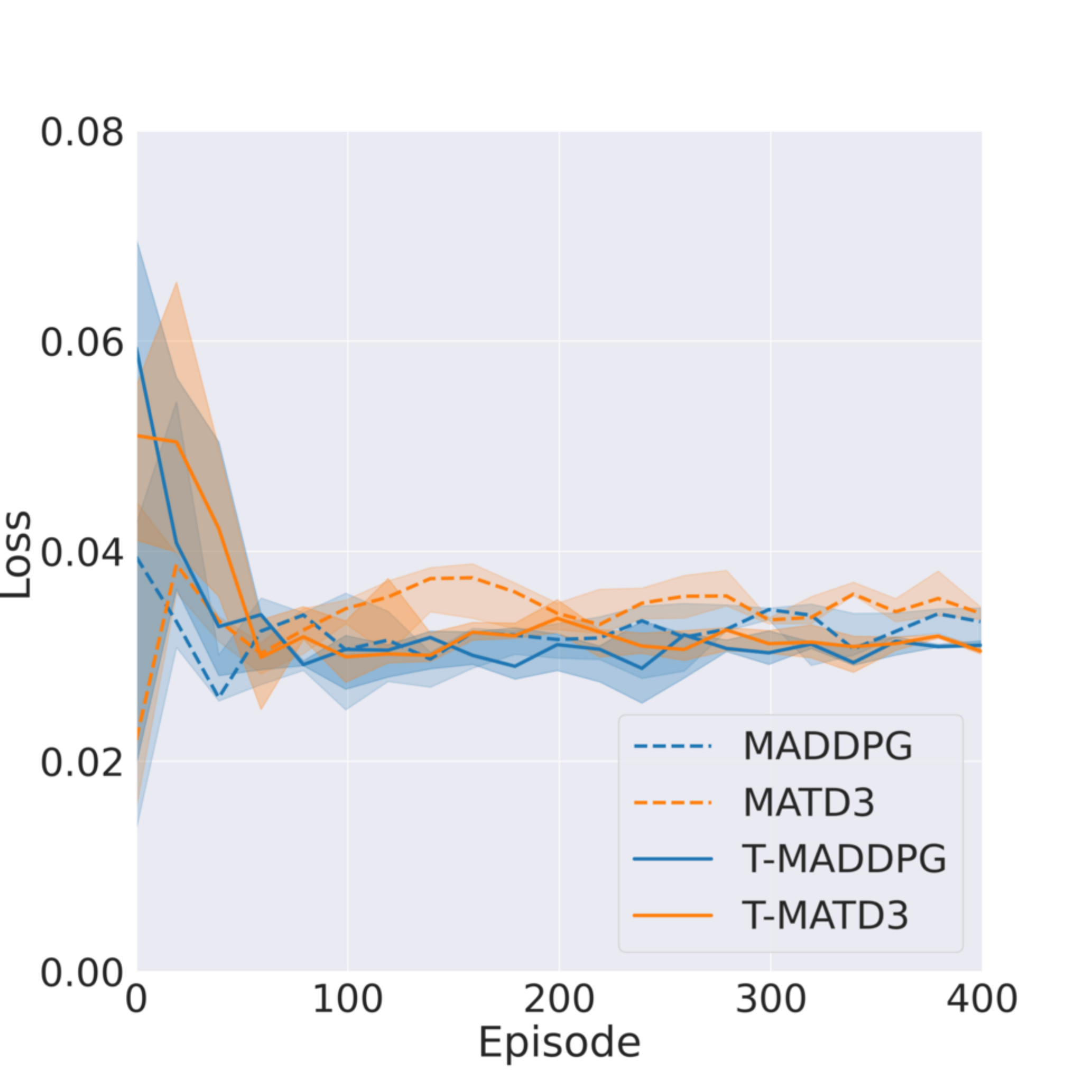}
    }
    \subfigure[QL-L2-322]{
        \includegraphics[scale=0.12]{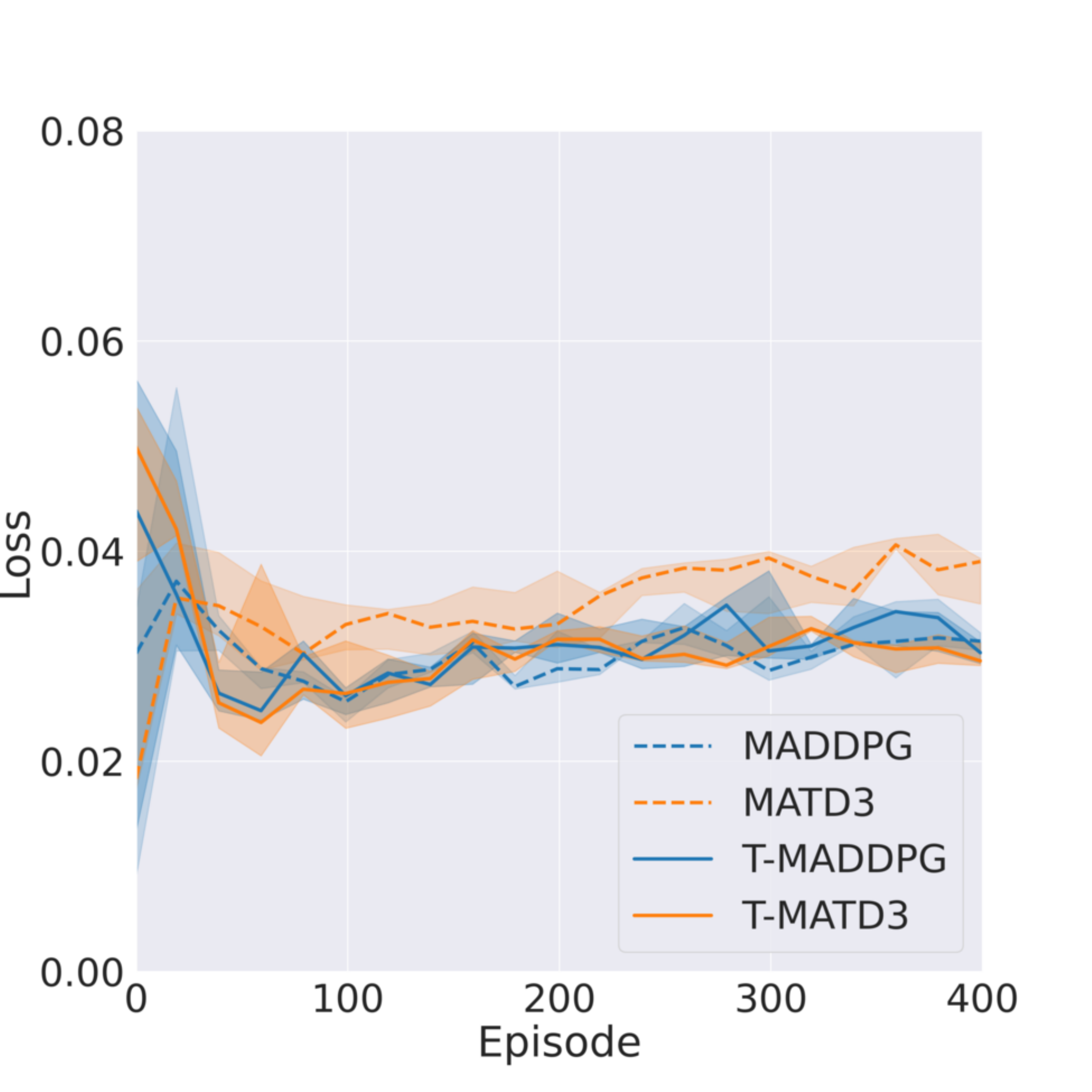}
    }
    \subfigure[QL-BL-322]{
        \includegraphics[scale=0.12]{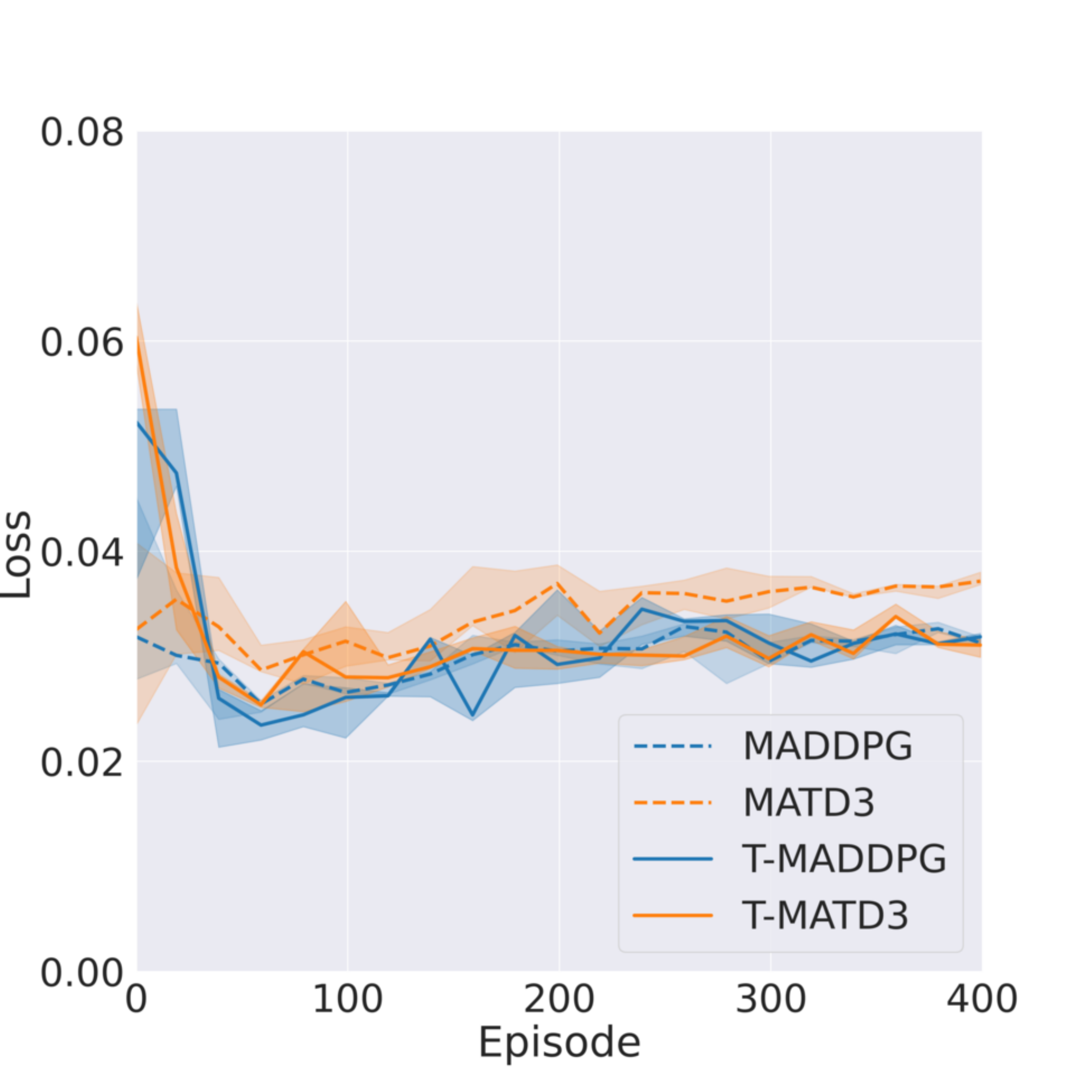}
    }
    \caption{Median CR and QL of algorithms with different voltage barrier functions. "T-" indicates the combination of T-MAAC and the baseline algorithm. The sub-caption indicates metric-Barrier-scenario and BL is the abbreviation of bowl.}
    \label{fig:main_result}
\end{figure*}

\begin{table*}%[htbp]
\caption{The mean test results in the test dataset. CR denotes the control rate; QL denotes the Q loss; PL denotes the power loss.}
\label{tab:performance}
\setlength{\tabcolsep}{0.7mm}
{
\begin{tabular}{l | ccc | ccc | ccc | ccc}
\toprule
\multirow{2}*{\textbf{Method}} & \multicolumn{3}{c}{\textbf{Spring}} & \multicolumn{3}{c}{\textbf{Summer}} & \multicolumn{3}{c}{\textbf{Fall}} & \multicolumn{3}{c}{\textbf{Winter}} \\
\cmidrule(lr){2-4}\cmidrule(lr){5-7}\cmidrule(lr){8-10}\cmidrule(lr){11-13}
& CR ($\%$) & QL ($\frac{\text{MW}}{\text{MVAR}}$) & PL (MW) & CR ($\%$) & QL ($\frac{\text{MW}}{\text{MVAR}}$) & PL (MW) & CR ($\%$) & QL ($\frac{\text{MW}}{\text{MVAR}}$) & PL (MW) & CR ($\%$) & QL ($\frac{\text{MW}}{\text{MVAR}}$) & PL (MW)\\
\midrule

322-MADDPG & 79.2 & 0.031 & 0.036 & 75.0 & 0.031 & 0.040 & 92.5 & 0.031 & 0.029  & 97.8 & 0.031 & 0.027\\
322-T-MADDPG & 88.6 & 0.027 & 0.037 & 86.3 & 0.027 & 0.044 & 96.7 & 0.025 & 0.026 & 98.0 & 0.024 & 0.021 \\
322-MATD3  & 59.4 & 0.033 & 0.037 & 54.1 & 0.033 & 0.040 & 77.9 & 0.034 & 0.033 & 89.4 & 0.035 & 0.033 \\
322-T-MATD3 & 90.6	& 0.028 & 0.039 & 89.5	& 0.029 & 0.046 & 97.9 & 0.028 & 0.028  & 99.3 & 0.027 & 0.024 \\
\midrule
141-MADDPG & 75.8 & 1.88 & 0.78 & 72.9	& 1.85 & 0.91 & 75.4 & 	1.95 & 0.53 & 71.1 & 1.99 & 0.45 \\
141-T-MADDPG  & 97.8 & 0.77 & 0.93 & 97.8 & 0.78 & 1.10 & 100 & 0.79 & 0.61 & 100 & 0.85 & 0.50\\
141-MATD3 & 78.7 & 1.83 & 0.93 & 73.4 & 1.77 & 1.08 & 90.1 & 1.95 & 0.67 & 89.9 & 2.03 & 0.59 \\
141-T-MATD3 & 97.4 & 0.77 & 0.89 & 97.4 & 0.79 & 1.06 & 99.9 & 0.79 & 0.57 & 100 & 0.85 & 0.47 \\
\bottomrule
\end{tabular}
}
\end{table*}

\section{Experiments}
\label{sec:experiments}
In this section, we conduct a series of experiments based on the MAPDN environment\cite{MAPDN} to evaluate the performance of T-MAAC.
We first introduce the experiment setups and implementation details of the algorithms.
Then, we compare the evaluation results of our algorithm with the baseline methods and the ablated variants. Furthermore, we show a case study that visualizes the attention weights in the self-attention mechanism to analyze which nodes in the distribution network agents should focus on (see Appendix \ref{sec:app-casestudy}).
\subsection{Experiment Setups}
\label{sec:setup}
The MAPDN\cite{MAPDN} is an environment of distributed/decentralized active voltage control on power distribution networks, which supports numerical studies for the 33-bus, 141-bus, and 322-bus network.
We conduct experiments in the 141-bus network scenario with 22 agents and the 322-bus network scenario with 38 agents. 

In order to evaluate different algorithms fairly, we randomly select some test scenarios from different seasons to construct a test dataset and a validation dataset(details in Appendix \ref{sec:app-setup}).
Each experiment is run with 5 random seeds and the test results during training are given by the median and the $25\%$-$75\%$ quartile shading.
And each experiment is evaluated in the validation dataset every 20 episodes during training.
After the training phase, we evaluate the learned strategy on the whole test dataset. 

Following the proposal of \cite{MAPDN}, we conduct main experiments with three different voltage barrier functions (see Appendix \ref{sec:function}) on the 141-bus and the 322-bus networks.
In experiments, we use two metrics to evaluate the performance of algorithms. \textit{Controllable rate (CR)}: It calculates the ratio of all buses' voltage being under control within safety range. \textit{Q loss (QL)}: It calculates the mean reactive power generations by agents per time step, which is the same as $l_q(\cdot)$ defined in Eq.(\ref{equ:reward_function}). \textit{QL} is an alternative metric to power loss because the power loss of the whole grid is hard to obtain in the actual distribution network.
CR is the most critical metric for the active voltage control task, and a low CR indicates that the entire grid is currently perilous.

\subsection{Baseline Methods and Implementation Details}
According to \cite{MAPDN}, MADDPG\cite{MADDPG} and MATD3\cite{MATD3} achieve excellent performance in the MAPDN environment compared to other state-of-the-art MARL algorithms.
Thus, we separately couple our proposed T-MAAC with MADDPG and MATD3 to evaluate the performance of our framework:
\begin{itemize}[leftmargin=*,topsep=0pt,parsep=0pt]
\item \textbf{MADDPG and MATD3}. In our experiments, the MLP-based policy network in MADDPG and MATD3 consists of one hidden layer and a GRU layer. And the MLP-based critic network is constructed with one hidden layer. 
\item \textbf{T-MADDPG and T-MATD3}. Compared to the baseline algorithms above, T-MADDPG and T-MATD3 replace the MLP-based network architecture with the transformer-based network architecture proposed in Section \ref{sec:policy} and \ref{sec:critic}. Moreover, additional auxiliary loss is introduced during training (see Algorithm \ref{alg:alg1}). In the policy network, the transformer encoder is composed of a stack of 2 transformer layers, and the embedding aggregation module is an extra transformer layer. The architecture of GRU head remains the same as the baseline algorithms above.

\end{itemize}
% Follow by \cite{MAPDN}, the behaviour policy/value network are update every 60 time steps and the target policy/value network are update every 120 time steps. 
Following \cite{SAC}, all algorithms are trained with the normalized reward and the action bound enforcement trick. We perform gradient clipping with L1 norm and the clip bound is set to 1.
Moreover, the parameters in policy networks are shared among agents, and the agent ID is concatenated with observation to distinguish different agents.
The hyper-parameters of algorithms are shown in Table \ref{tab:commonhyper} (see Appendix \ref{sec:app-hyperparameters}).

\subsection{Result}

The median CR and QL of algorithms during training for MADDPG, MATD3, T-MADDPG and T-MATD3 are shown in Figure \ref{fig:main_result}.
As the figure shows, our proposed T-MAAC framework consistently improves the performance of baseline methods under three types of rewards and two different scale grid scenarios. 
Owing to the learned better representations relevant to the active voltage control task, T-MADDPG and T-MATD3 improve the controllable rate \textit{(CR)} in the grid while reducing the reactive power generation \textit{(QL)}.
T-MAAC significantly performs well on the 141-bus network, verifying the importance of better representations.
As for the 322-bus network, a large-scale scenario with 38 agents, T-MADDG increases the CR while maintaining the low QL as same as MADDPG, and T-MATD3 achieves a higher CR with a lower QL.
By comparing the performance with three different reward functions, T-MAAC also stabilizes the training process and alleviates the phenomenon mentioned in \cite{MAPDN} that algorithms is sensitive to reward functions.
% TODO
The better representations improve sample efficiency and resolve the issue that different reward functions lead agents to different local optimal policies. 

We also evaluate all algorithms trained by L2-shape voltage barrier function in the test dataset. We show the performances in Table \ref{tab:performance} (including actual power loss \textit{PL} in the whole distribution network).
The metric \textit{CR} shows that it is more challenging to control the voltage in spring and summer than in fall and winter due to excessive active power injection produced by PVs in the former.
Our proposed T-MAAC overcomes such difficulties and achieves better performances in all scenarios, especially in spring and summer.
Meanwhile, penalized by \textit{Q loss}, T-MAAC learns a strategy with less reactive power generation than baseline methods. It is worth noting that \textit{QL} is a proxy for power loss during training, thus, the learned strategy with less \textit{QL} may still lead to more \textit{PL}. Such a matter can be alleviated by more related and easy-to-obtain surrogate metric\cite{AVC_SAC}.

\subsection{Ablation Study}

In this section, we conduct a series of experiments to examine further which particular components of T-MAAC are essential for the performance. We fix the baseline algorithm to be MADDPG trained by L2-shape voltage barrier function.
The performances of the variants of T-MADDPG are evaluated in the following experimental settings.

\subsubsection{Auxiliary Task}

\begin{figure}
    \centering
    \subfigure[CR-L2-141]{
        \includegraphics[scale=0.12]{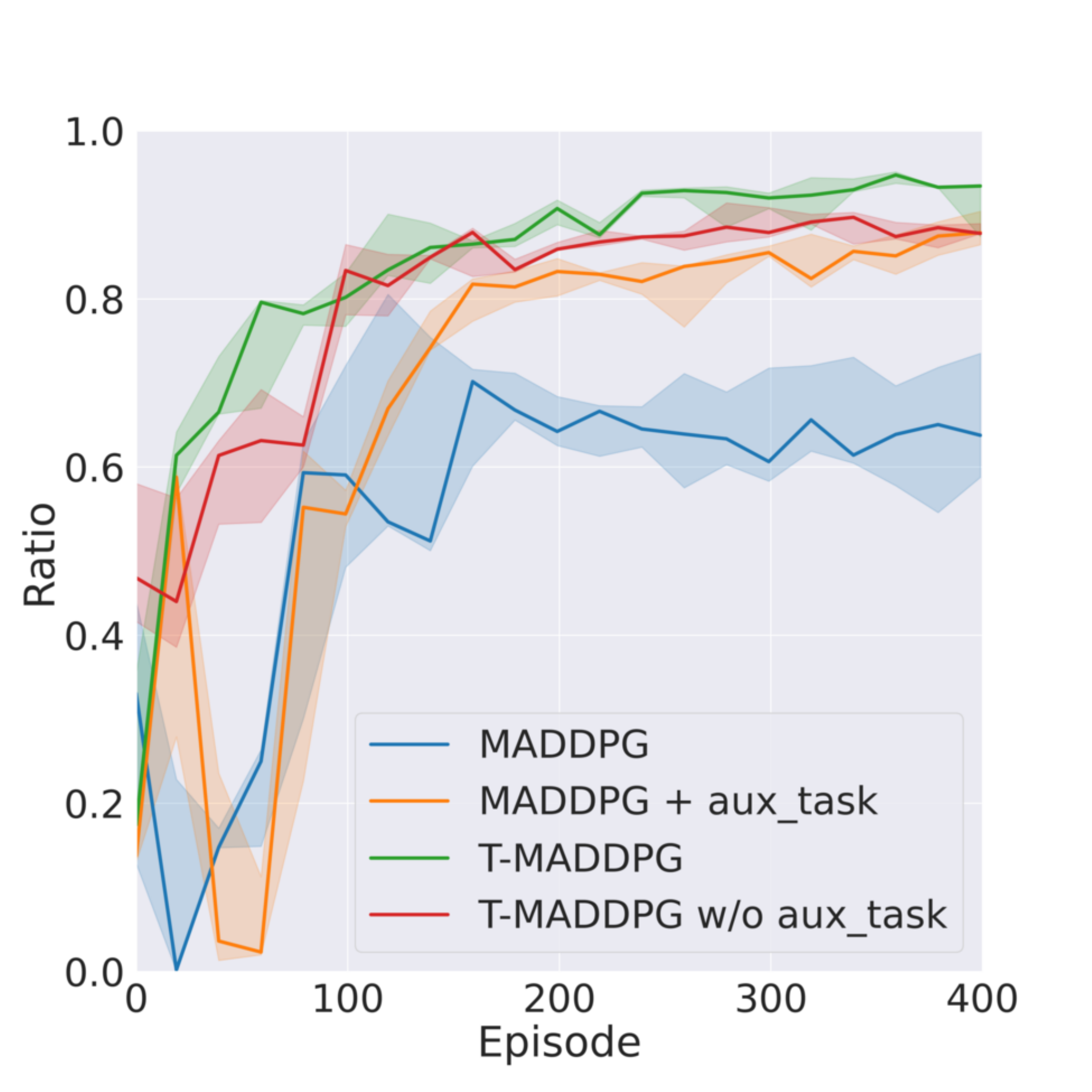}
    }\hspace{8mm}
    \subfigure[CR-L2-322]{
        \includegraphics[scale=0.12]{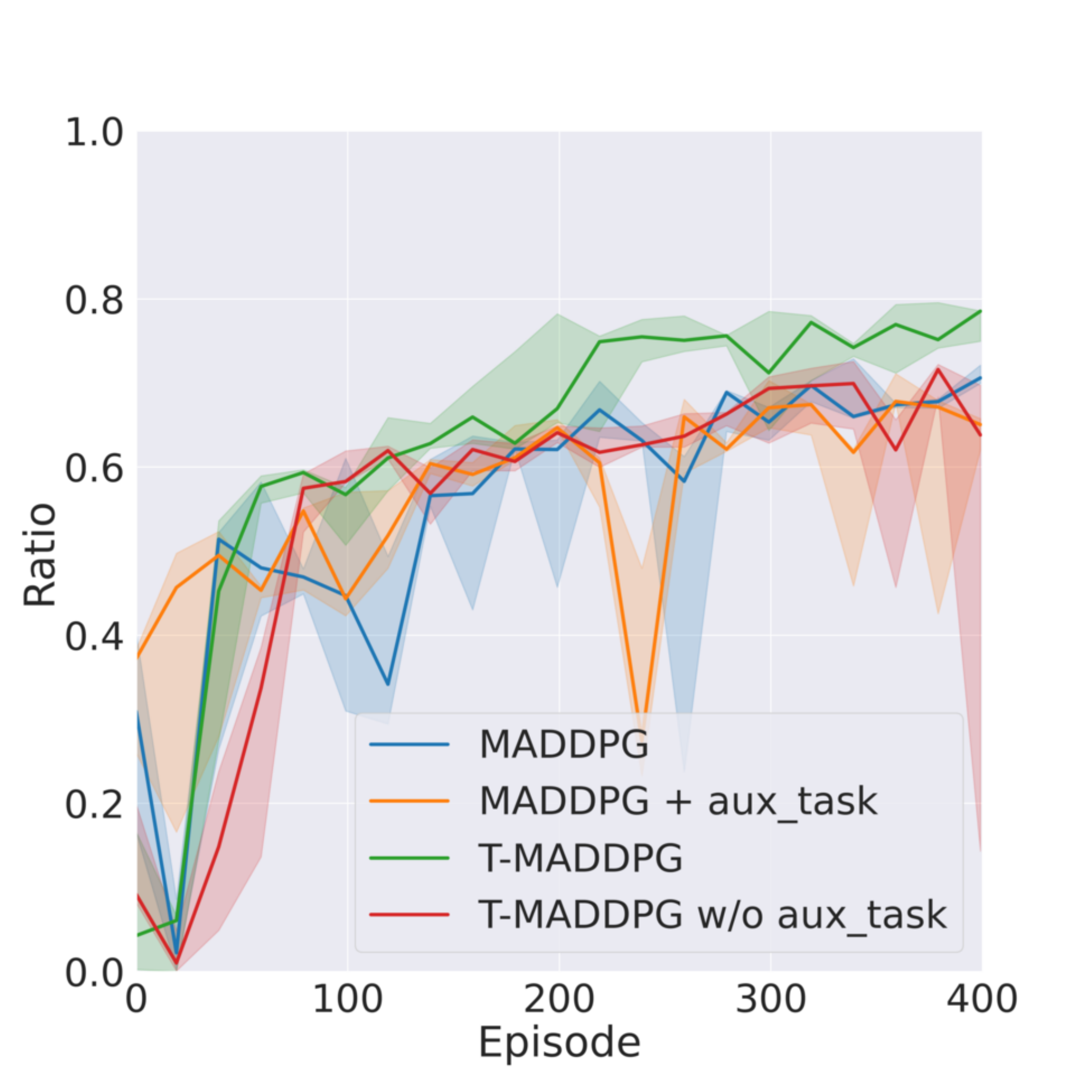}
    }\vspace{-4mm}
    \\
    \subfigure[QL-L2-141]{
        \includegraphics[scale=0.12]{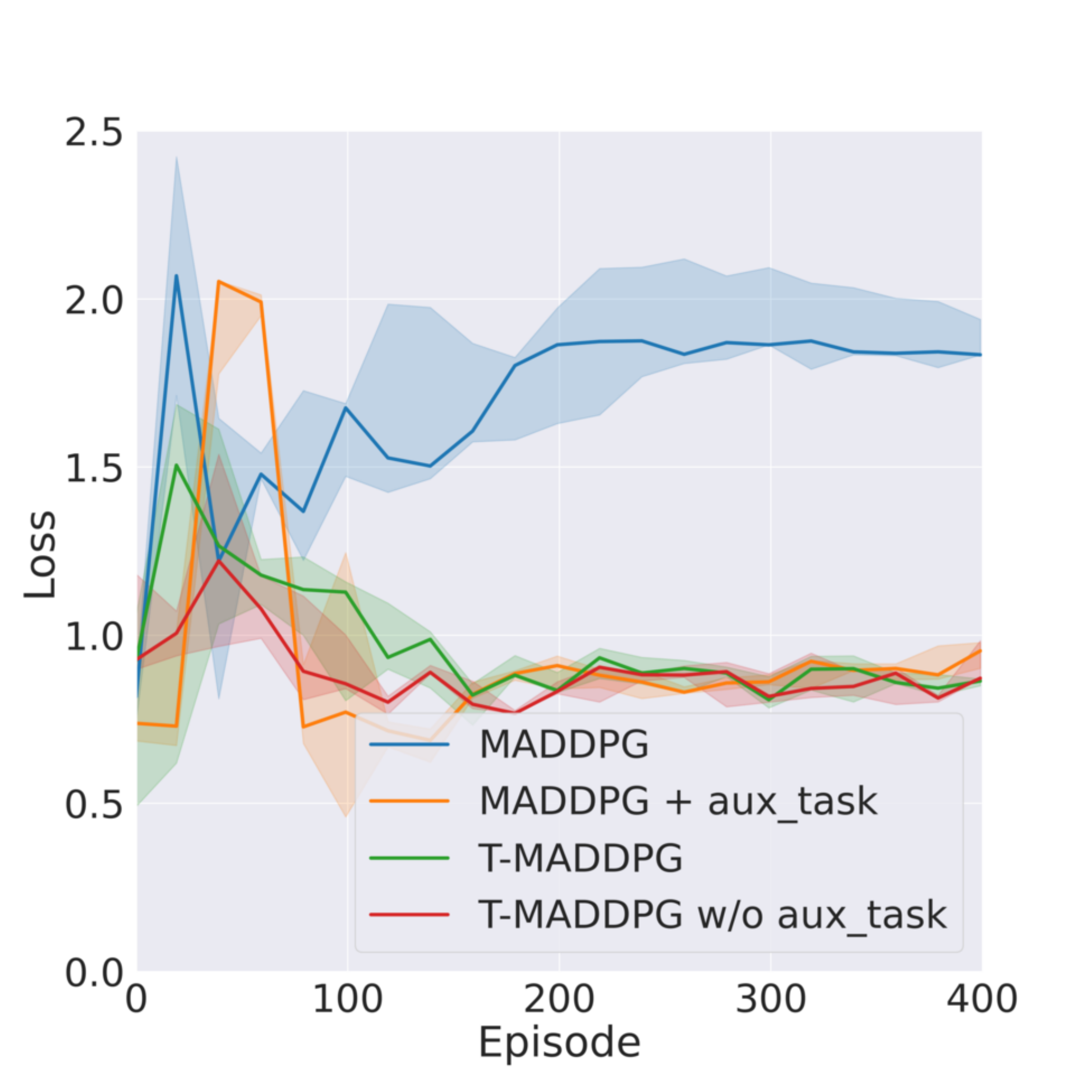}
    }\hspace{8mm}
    \subfigure[QL-L2-322]{
        \includegraphics[scale=0.12]{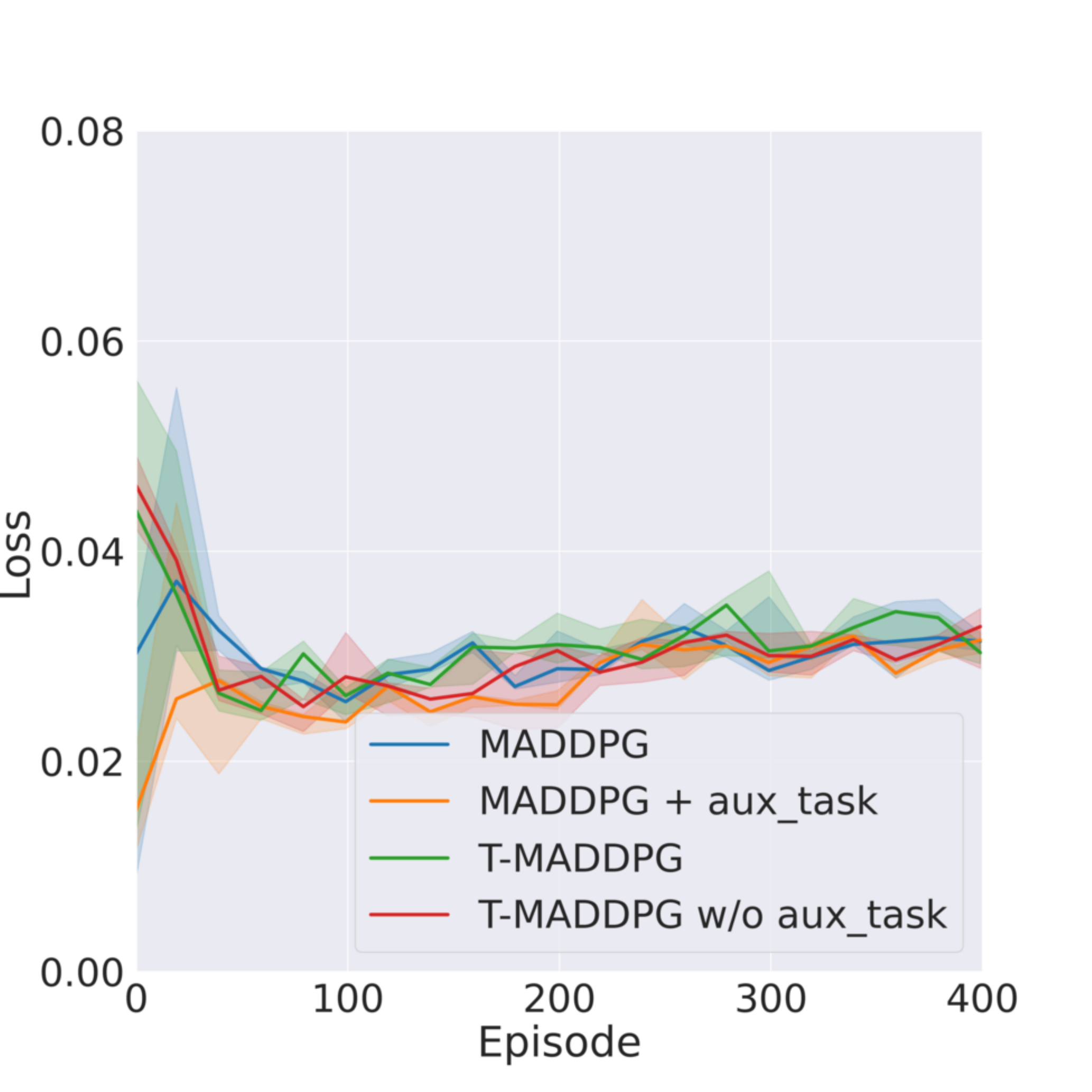}
    }
    \caption{Performance comparison with or w/o auxiliary task. The sub-caption indicates metric-Barrier-scenario.}
    \label{fig:compare_aux}
\end{figure}

% Transformer architectures have achieved breakthrough success in NLP and CV due to their ability both effectively aggregate information over long time horizons and scale to a large amount of data.
% However, transformers have not yet been fully explored in reinforcement learning domains, because that the standard transformer architecture is difficult to optimize\cite{stabilize_transformer}.
% We also find that optimizing the policy networks by Eq.~\eqref{equ:actor_loss} cannot lead to performance gain in our experiments.

We add an additional auxiliary task during training to stabilize the training process (see Section \ref{sec:auxiliary}).
The following ablated variants are designed to verify the effectiveness of the auxiliary task:
\begin{itemize}[leftmargin=*,topsep=0pt,parsep=0pt]
\item \textbf{T-MADDPG w/o aux task}: We remove the auxiliary task and optimize model by Eq.~\eqref{equ:actor_loss} and Eq.~\eqref{equ:critic_loss}.
\item \textbf{MADDPG with aux task}: We also add the auxiliary task to the MLP-based MADDPG to figure out whether the auxiliary task can improve performance with different network architectures.
\end{itemize}
The result is shown in Figure \ref{fig:compare_aux}.
The performance of T-MADDPG is better than T-MADDPG without auxiliary task during almost the entire training process especially on the 322-bus network that is a more challenging large-scale scenario.
T-MADDPG without auxiliary task performs little differently from MADDPG, which indicates training with auxiliary task is essential for our transformer-based network architectures.

In addition, training with auxiliary task also improves the performance of MADDPG on the 141-bus network, which means that the auxiliary loss during training helps convergence to a better policy.
However, training with auxiliary task doesn't work on the 322-bus network.
This may be due to the fact that the simple policy/critic network constructed with MLPs is no longer able to capture the nature in the grid, especially in the large-scale grid.

% \subsubsection{Embedding Aggregation Module in Policy Network}
\subsubsection{Ablation Study on the Transformer-based Policy Network}

\begin{figure}
    \centering
    \subfigure[CR-L2-141]{
        \includegraphics[scale=0.12]{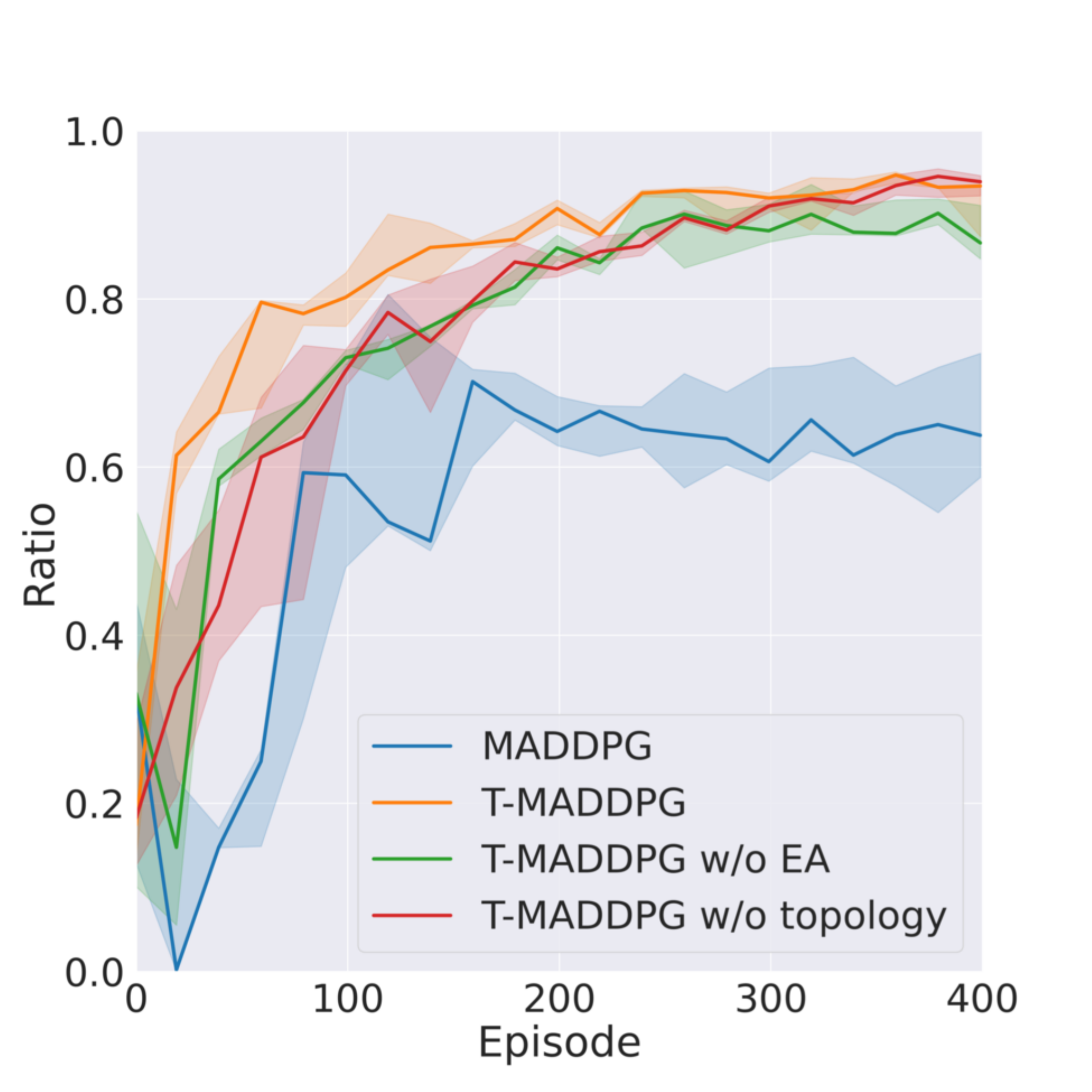}
    }\hspace{8mm}
    \subfigure[CR-L2-322]{
        \includegraphics[scale=0.12]{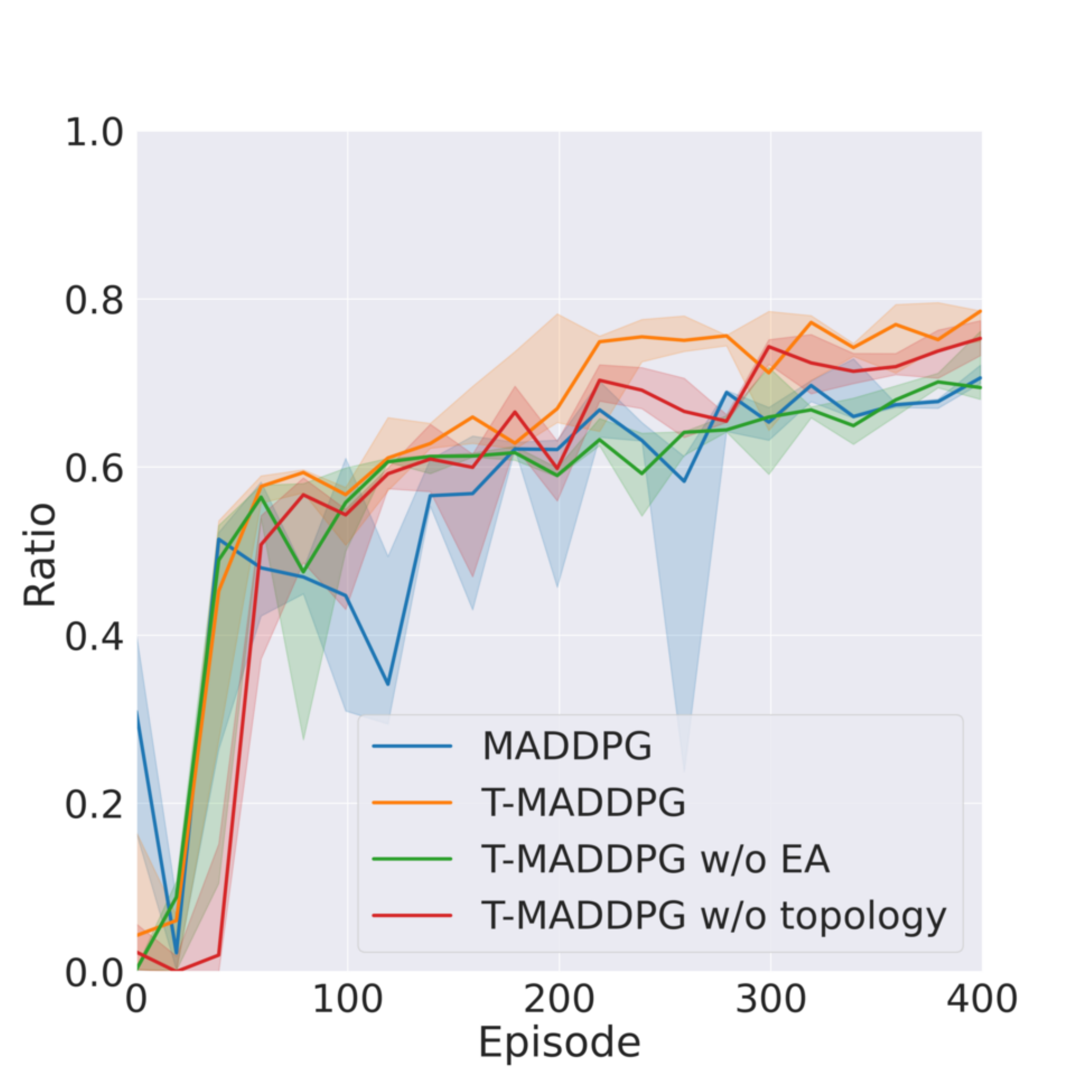}
    }\vspace{-4mm}
    \\
    \subfigure[QL-L2-141]{
        \includegraphics[scale=0.12]{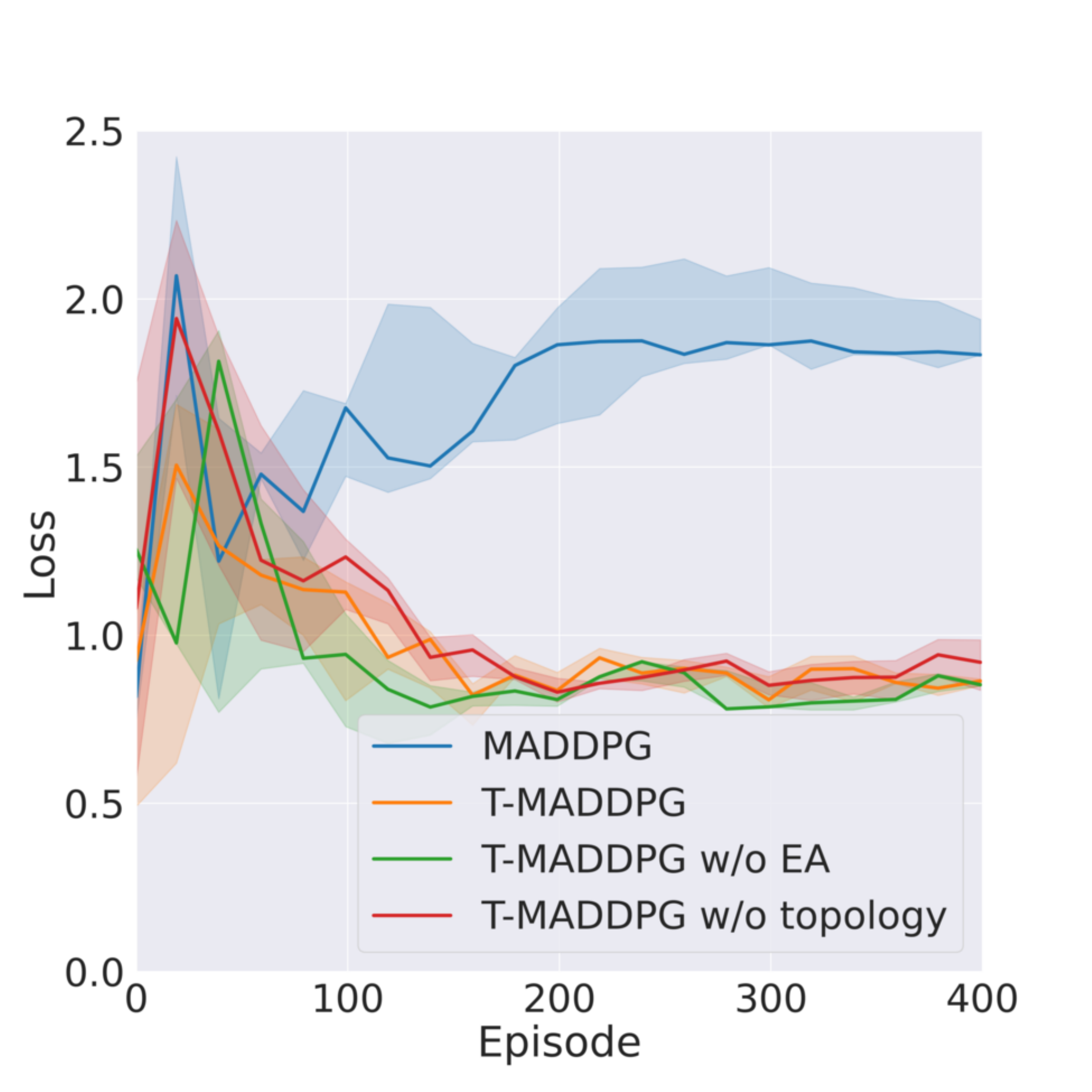}
    }\hspace{8mm}
    \subfigure[QL-L2-322]{
        \includegraphics[scale=0.12]{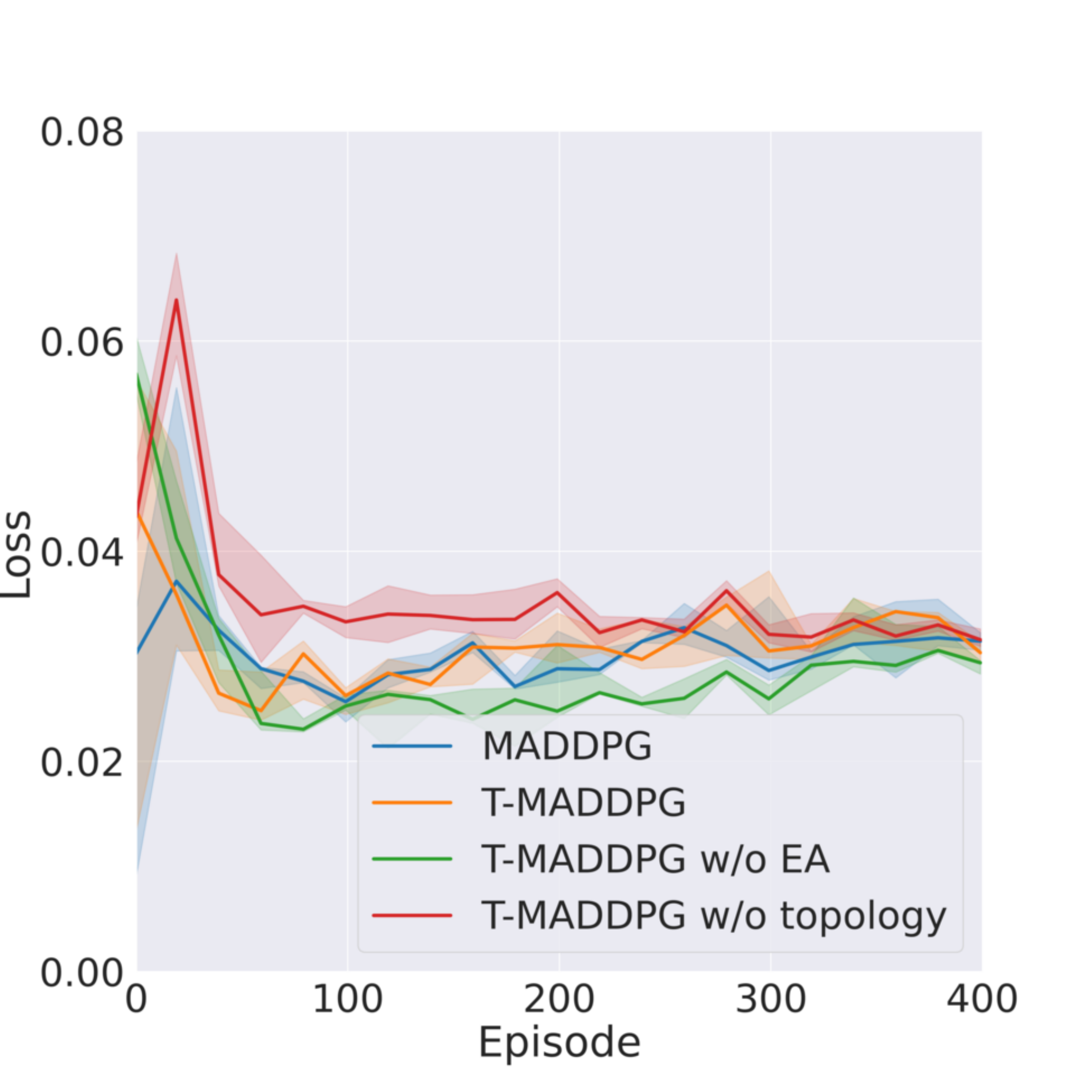}
    }
    \caption{Performance comparison on variants of our policy network. The sub-caption indicates metric-Barrier-scenario.}
    \label{fig:compare_arch}
\end{figure}
In T-MAAC, we design a novel policy network based on transformer to achieve better performance on active voltage control task (see Section \ref{sec:policy}).
We introduce two ablated variants to examine modules in the transformer-based policy network as follows:
\begin{itemize}[leftmargin=*,topsep=0pt,parsep=0pt]
\item \textbf{T-MADDPG w/o EA}: We remove the embedding aggregation module, and the final representation of $\mathcal{O}$ is obtained by average outputs of the transformer encoder.
\item \textbf{T-MADDPG w/o topology}: We remove the adjacency matrix and the mask in the self-attention mechanism. Attention operations in Eq.~\eqref{equ:self-attention} are implemented between all nodes regardless of whether they are connected in the zone or not.
\end{itemize}

The embedding aggregation module integrate node-based information to global information from the point of the agent, and the result in Figure \ref{fig:compare_arch} shows that it further improves performance on both 141-bus network and 322-bus network.
Further more, by comparing the performance of T-MADDPG w/o EA and MADDPG in the 322-bus network, it can be seen that vanilla transformer architecture without EA can't handle various observation space. This result verifies the opinions discussed in Section \ref{sec:policy} that aggregating information from the perspective of decision maker allows transformer encoder to obtain better representations suitable for this task.

If we don't inject the position information into the transformer, all nodes are treated equally in the early stages of training.
Thus, inspired by \cite{stabilize_transformer,SMAAC}, we select the adjacency matrix as the mask in the self-attention mechanism to assist agents in capturing the nature of the grid.
As shown in Figure \ref{fig:compare_arch}, the utilization of topology improves sample efficiency in the baseline algorithm.

\subsubsection{Transformer-based Critic}

\begin{figure}
    \centering
    \subfigure[CR-L2-141]{
        \includegraphics[scale=0.12]{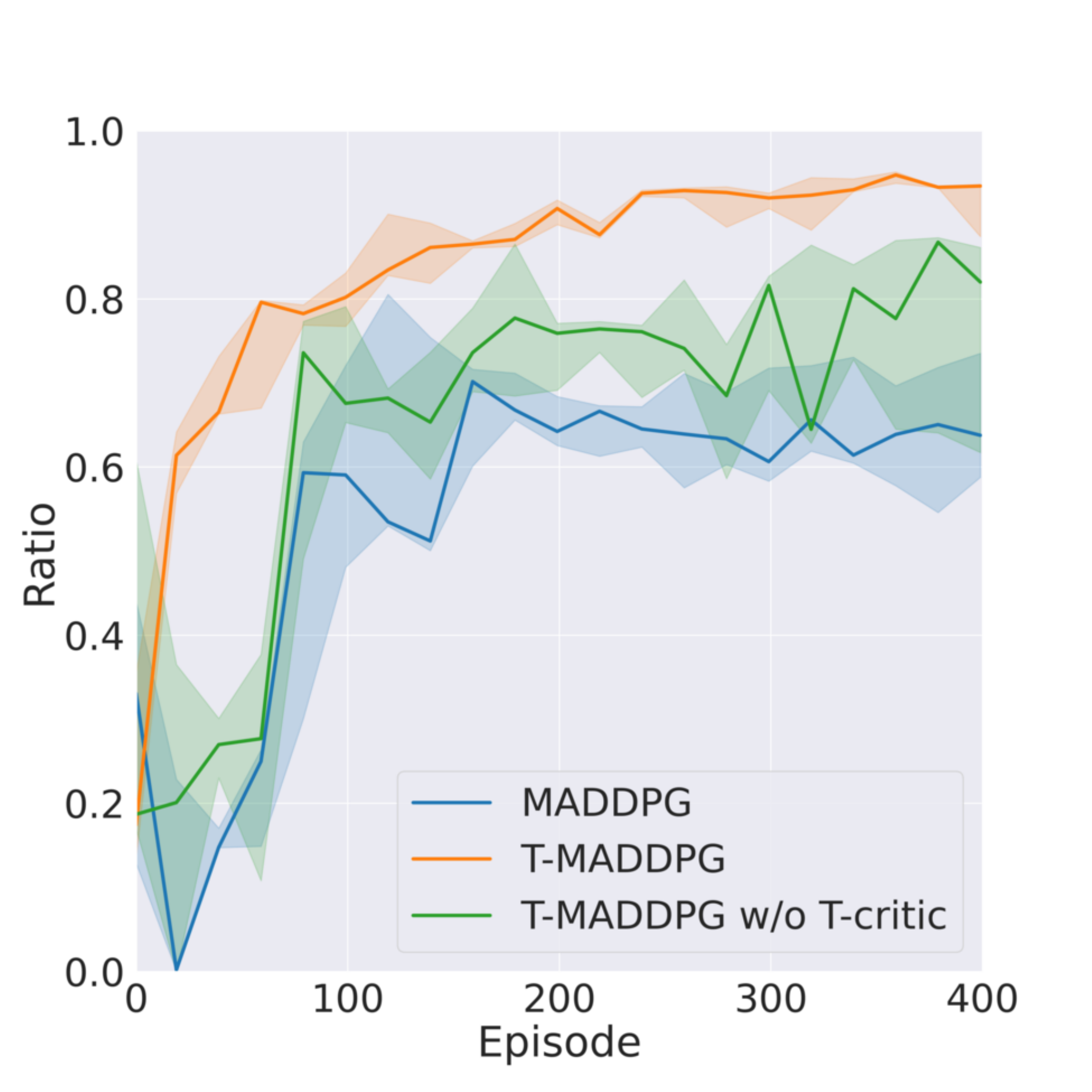}
    }\hspace{8mm}
    \subfigure[CR-L2-322]{
        \label{fig:ablation_critic_141}
        \includegraphics[scale=0.12]{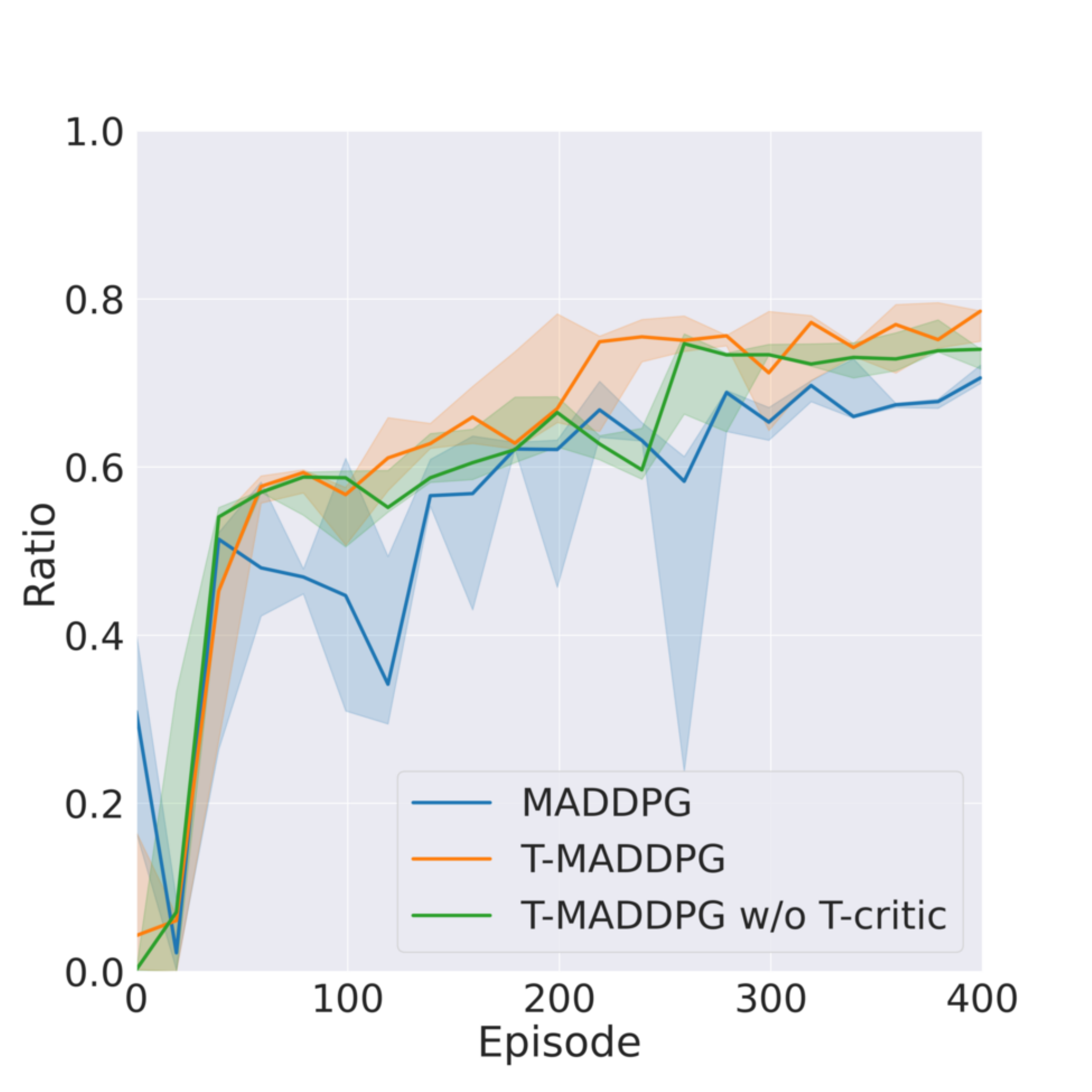}
    }\vspace{-4mm}
    \\
    \subfigure[QL-L2-141]{
        \includegraphics[scale=0.12]{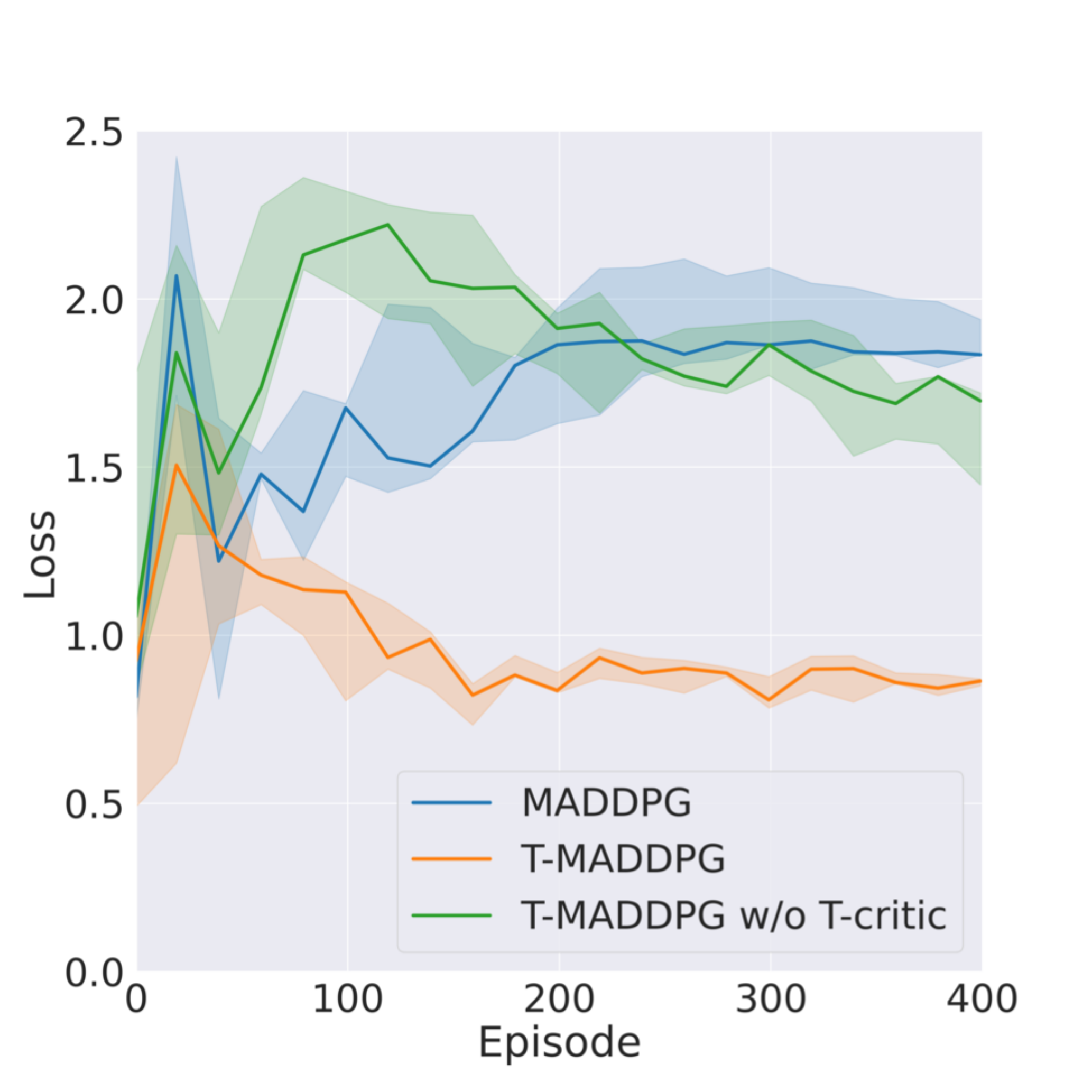}
    }\hspace{8mm}
    \subfigure[QL-L2-322]{
        \includegraphics[scale=0.12]{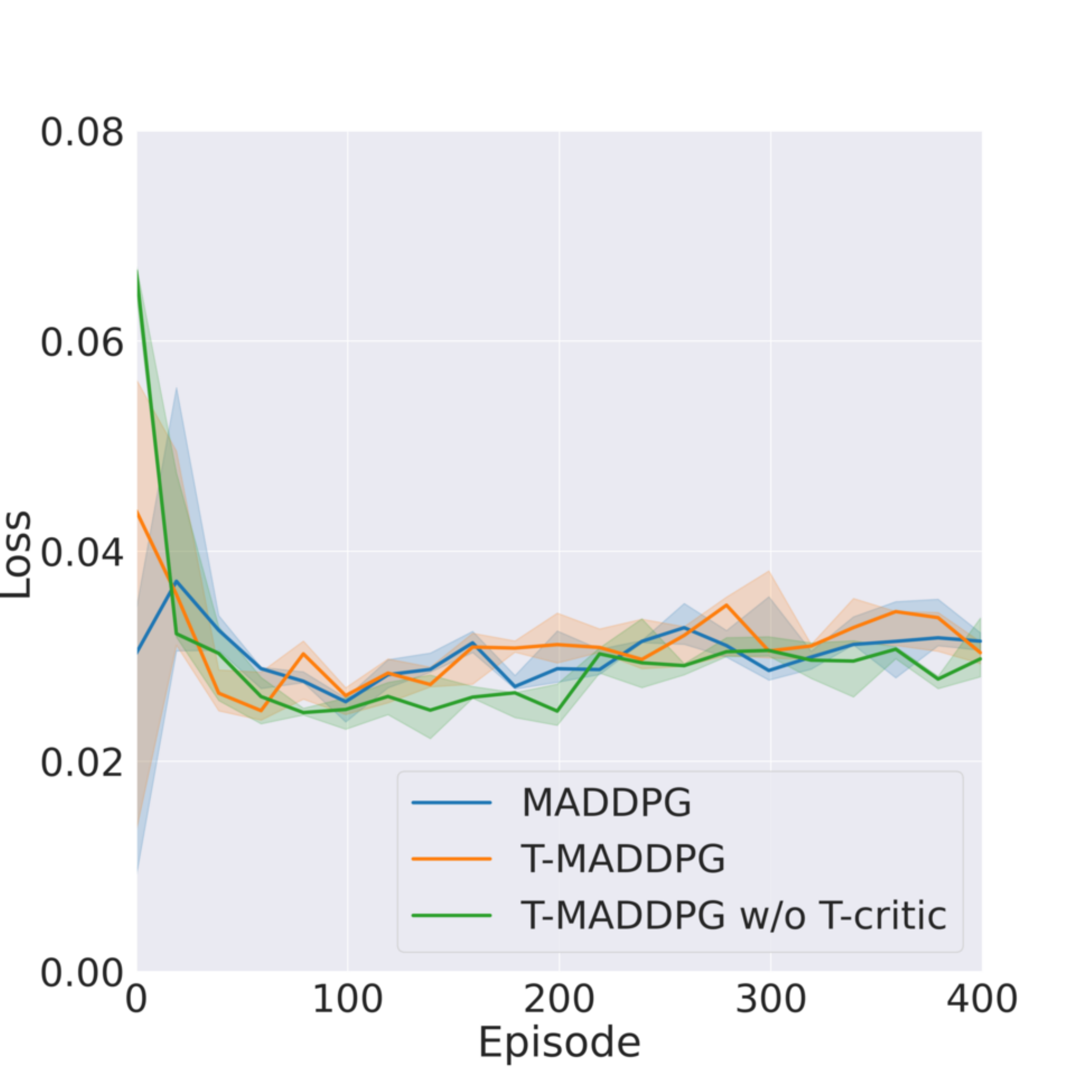}
    }
    \caption{Performance comparison on variants of our critic network. The sub-caption indicates metric-Barrier-scenario.}
    \label{fig:compare_critic}
\end{figure}

In this section, we conduct ablation experiments to figure out the effect of introducing transformers into the global critic network.
\begin{itemize}[leftmargin=*]
\item \textbf{T-MADDPG w/o T-critic}: We replace the transformer-based critic network with the widely used MLP-based critic network as same as MADDPG.
\end{itemize}
We show the result in Figure \ref{fig:compare_critic}.
Compared to T-MADDPG w/o transformer-based critic, T-MADDPG achieves better performance on both scenarios.
Moreover, the transformer-based critic also stabilizes the training process in 141-bus network as shown in Figure \ref{fig:ablation_critic_141}.

% The performance improvement achieved by transformer-based critic network can be credited to the self-attention mechanism. Here, we use the 141-bus network as an example to demonstrate how attention mechanism benefits the cooperation between agents during training.

\section{Conclusions}
\label{sec:conclusions}
In this paper, we propose T-MAAC, a transformer-based multi-agent actor-critic framework, for voltage stabilization in power distribution networks. Our framework consists of a policy network and a global critic network. 
The policy network based on transformer captures the characteristics of grid, obtaining better representations for power network task.
In the global critic network, we introduce the self-attention mechanism to model the correlation between agents and achieving better performance.
Additionally, we adopt the auxiliary task, predicting the voltage out of control ratio in a zone for active voltage control task, to stabilize the training process and improve the embedding learning.
We conduct extensive evaluations as well as ablation studies in the real-world scale grid scenarios provided by MAPDN.
The experimental results demonstrate that T-MAAC significantly improves the performance of existing MARL algorithms for voltage stabilization.
\begin{acks}
This work was supported in part by the National Natural Science Foundation of China under Contract 61836011 and in part by the Youth Innovation Promotion Association CAS under Grant 2018497. It was also supported by the GPU cluster built by MCC Lab of Information Science and Technology Institution, USTC.
\end{acks}


\bibliographystyle{ACM-Reference-Format}
\bibliography{KDD22}

\clearpage
\appendix

\section{Experimental Settings}

\subsection{MAPDN Environment and Datasets}
\label{sec:app-setup}
In MAPDN, the 33-bus and 141-bus networks are modified from IEEE 33-bus\cite{IEEE33} and IEEE 141-bus\cite{IEEE141}, respectively, while the 322-bus network is constructed by topology from SimBench\cite{SimBench}.
The data supporting simulation (i.e., load profile, load and PV data, active and reactive power consumption) are collected from the real world and then interpolated with the 3-min resolution consistent with the grid's real-time control period.
To guarantee the safety of distribution network, the environment manually sets the range of actions with $[-0.6, 0.6]$ for the 141-bus case and $[-0.8,0.8]$ for the 322-bus case suggested by MAPDN\cite{MAPDN}.
As for the reward function, $\alpha$ in Eq.(\ref{equ:reward_function}) is set to 0.1 in the 322-bus case and 0.01 in the 141-bus case to tune the trade-off between voltage control performance and reactive power generation loss.

It is worth noting that the difficulty of voltage control problem varies during different months of a year. For example, during the midday summer, excessive active power from intense sunlight is injected into the grid, creating a more significant challenge for the voltage control task than in winter.
Thus, a series of fixed scenarios must be chosen to evaluate algorithms fairly.
We randomly select 10 episodes per month, a total of 120 episodes, which constitute the test dataset.
Each episode lasts for 480 time steps (i.e., a day).
Then, we split the test dataset into four parts by month: Spring (Mar., Apr., May.), Summer (Jun., Jul., Aug.), Fall (Sept., Oct., Nov.), Winter (Dec., Jan., Feb).
A validation dataset is obtained by randomly selecting 10 episodes from the test dataset.
During the training phase, we randomly sample the initial state for an episode and each episode lasts for 240 time steps (i.e., a half day).
Each experiment is run with 5 random seeds and the test results during training are given by the median and the $25\%$-$75\%$ quartile shading as same as MAPDN\cite{MAPDN}.
Moreover, each experiment is evaluated in the validation dataset every 20 episodes during training.
After the training phase, we evaluate the learned strategy on the whole test dataset. 

\subsection{Network Topology}
\label{sec:network_topology}
\begin{figure}[h]
    \centering
    \subfigure[141-bus network]{
        \includegraphics[scale=0.20]{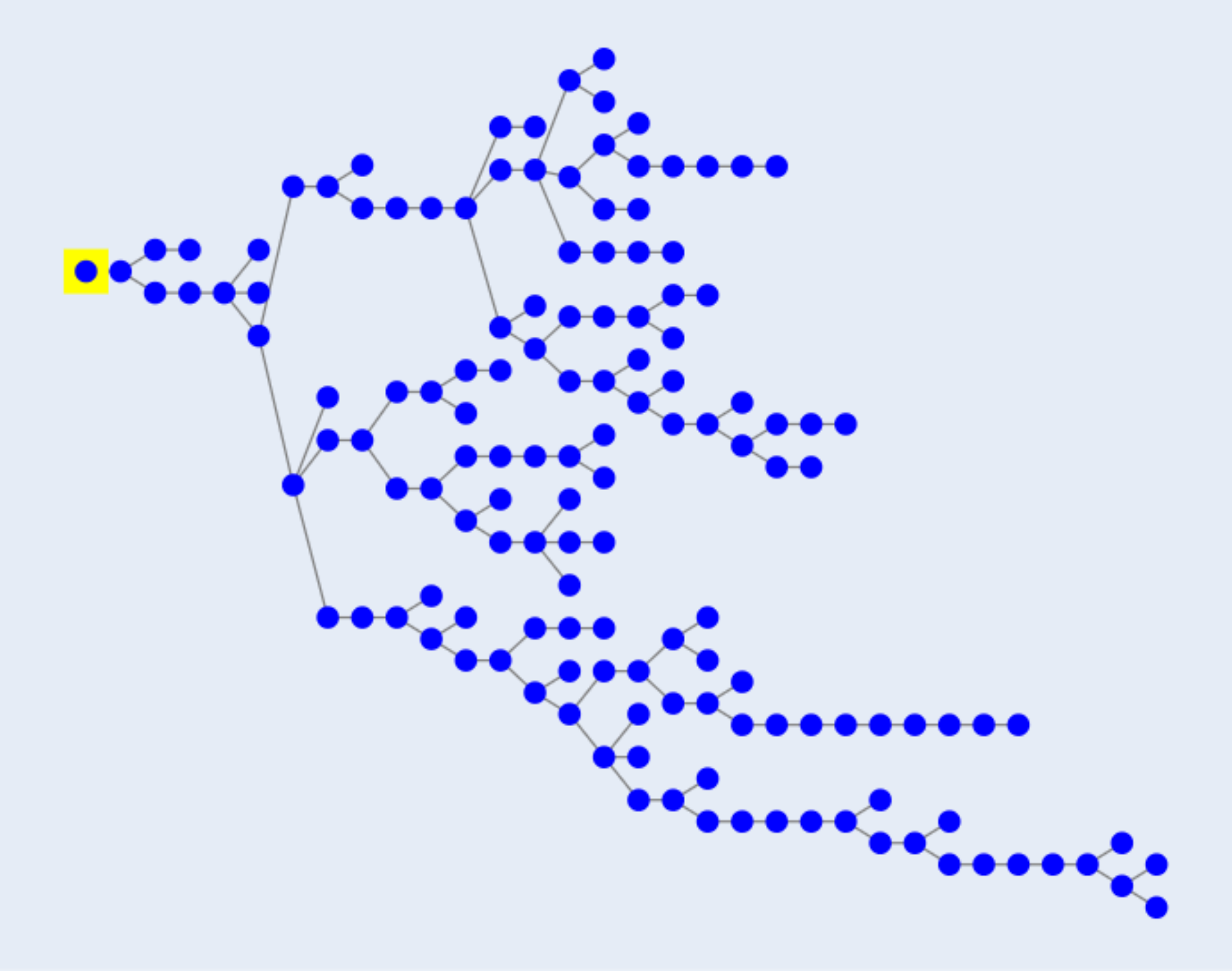}
    }
    \subfigure[322-bus network]{
        \includegraphics[scale=0.20]{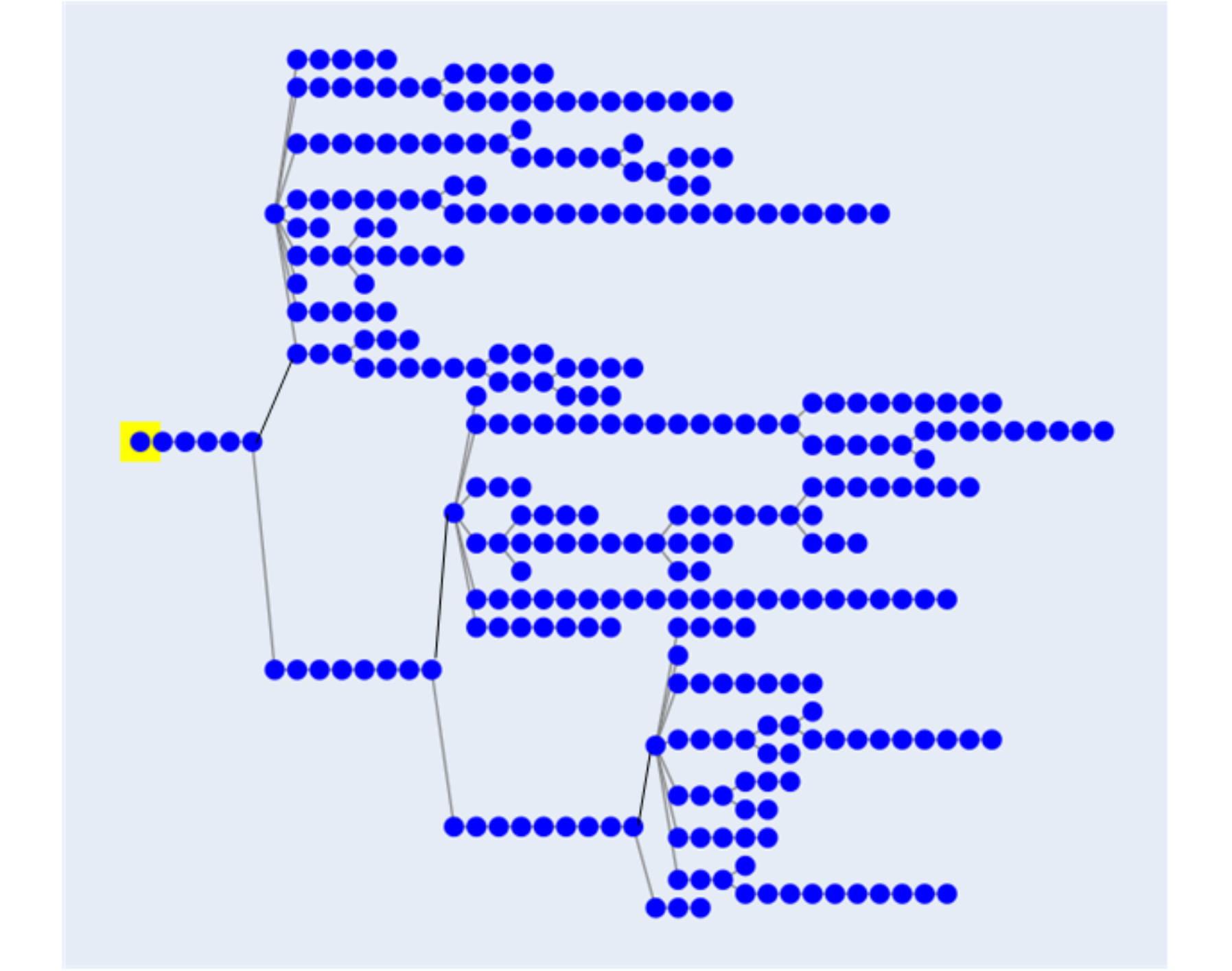}
    }
    \caption{The topologies of power networks. Figures are visualized by the PandaPower toolkit\cite{pandapower}.}
    \label{fig:network_topology}
\end{figure}
The network topologies of the 141-bus network and the 322-bus network are shown in Figure \ref{fig:network_topology}.
In the 141-bus network, there are 84 loads connected to some specific nodes, and 22 PVs (agents) are installed in some specific nodes.
As for the 322-bus network, 337 loads and 38 PVs are connected to some specific nodes.

\subsection{Reward Functions}
\label{sec:function}
\begin{figure}[h]
  \centering
  \includegraphics[scale=0.3]{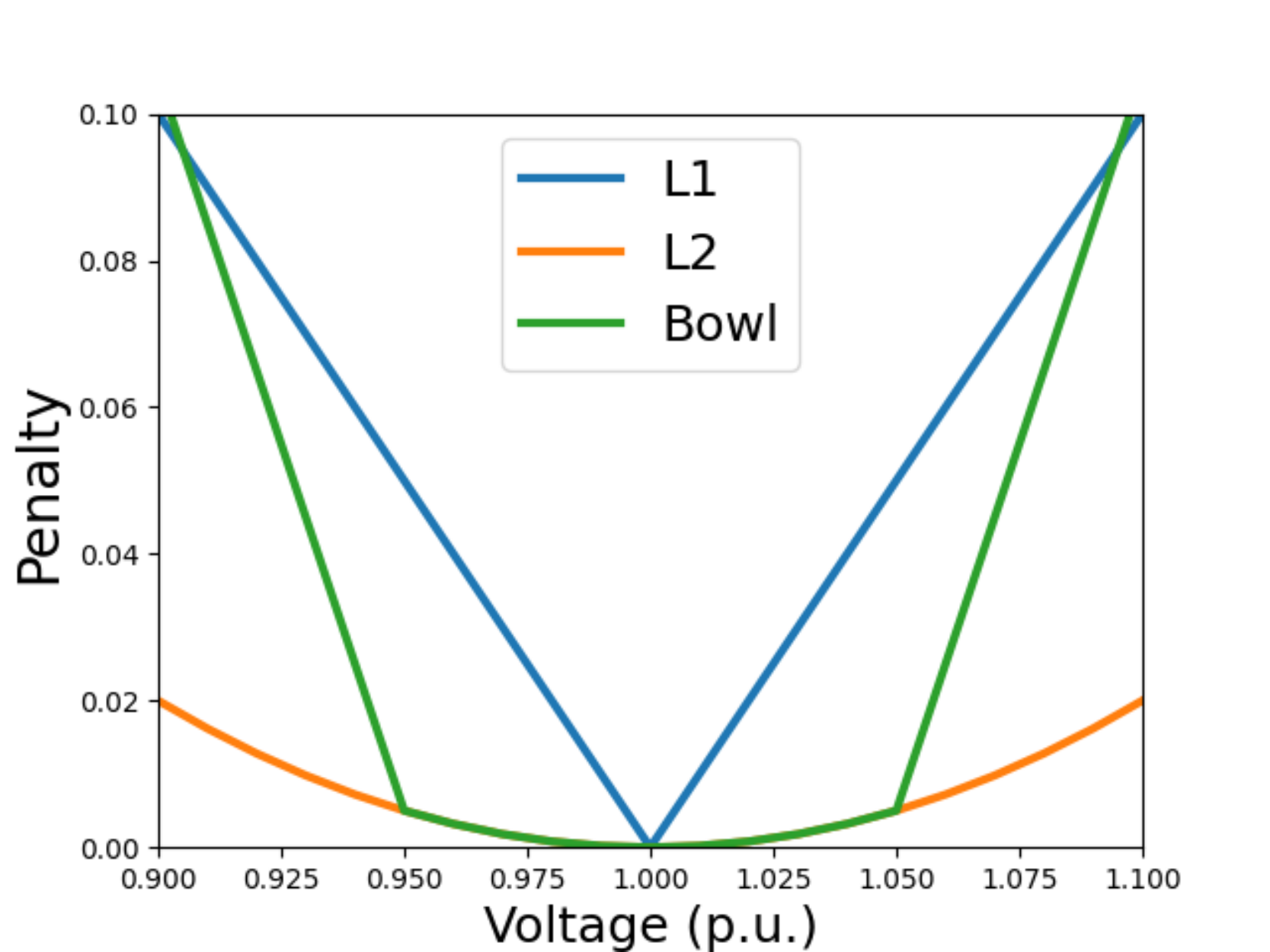}
  \caption{Three different voltage barrier functions proposed by \cite{MAPDN}.}
  \label{fig:function}
\end{figure}
In this work, reward functions are configured in accordance with the guidelines in MAPDN environment \cite{MAPDN}. We also conduct main experiments with different voltage barrier functions using the same settings as \cite{MAPDN}: L1-shape, L2-shape and Bowl-shape as shown in  Figure \ref{fig:function}.
Although the action range has been limited, there is still a possibility that the whole power system crashes due to incorrect control actions. To address this issue, if the power system crash, the system would backtrack to the last state and terminate the simulation with a reward of -200 regard as extra penalty.  

\subsection{Hyper-parameters of Algorithms.}
\label{sec:app-hyperparameters}
The hyper-parameters in our algorithms are shown in Table \ref{tab:commonhyper}.
\begin{table}[h]
  \caption{Hyper-parameters in experiments.}
  \label{tab:commonhyper}
  \begin{tabular}{l|l}
    \toprule
    \textbf{Name} & \textbf{Value}\\
    \midrule
    \textit{Common:} \\
    \quad optimizer & RMSProp\cite{RMSprop} \\
    \quad policy learning rate & $10^{-4}$ \\
    \quad value learning rate & $10^{-4}$ \\
    \quad policy update epochs & $1$ \\
    \quad value update epochs & $10$ \\
    \quad target update learning rate & $0.1$ \\
    \quad discount factor & $0.99$ \\
    \quad replay buffer size & $5000$ \\
    \quad batch size & $32$ \\
    \midrule
    \textit{MADDPG and MATD3:} \\
    \quad MLP hidden dimension (policy) & [64] \\
    \quad GRU hidden dimension (policy) & 64 \\
    \quad MLP hidden dimension (critic) & [64] \\
    \midrule
    \textit{T-MADDPG and T-MATD3:} \\
    \quad auxiliary task learning rate & $10^{-5}$ \\
    \quad auxiliary task update epochs & $10$ \\
    \quad transformer hidden dimension (policy) & $64$ \\
    \quad transformer layers (policy) & $3$ \\
    \quad transformer hidden dimension (critic) & $64$ \\
    \quad transformer layers (critic) & $3$ \\
    \quad number of multi-attention heads & $4$ \\
  \bottomrule
\end{tabular}
\end{table}

\clearpage
\section{Extra Experimental Results}
\subsection{Case Study}
\label{sec:app-casestudy}
\begin{figure}[h]
    \centering
    \subfigure[A zone in the 322-bus network. The sun emoji represents the location where a PV is installed. And the PV installed in node 15 is selected as a special agent to visualize attention weights.]{
        \label{fig:case-network}
        \includegraphics[scale=0.2]{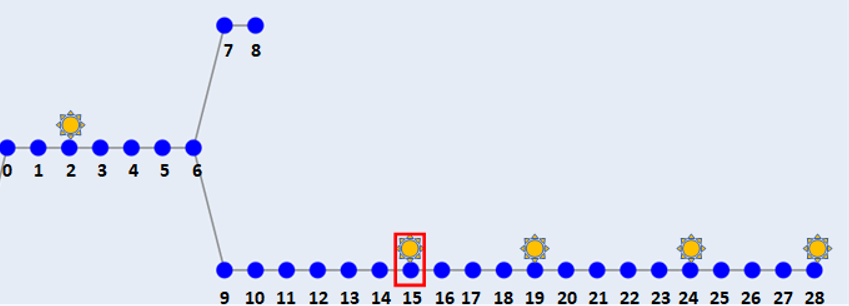}
    }
    \\
    \subfigure[Voltage]{
        \label{fig:case-voltage}
        \includegraphics[scale=0.1]{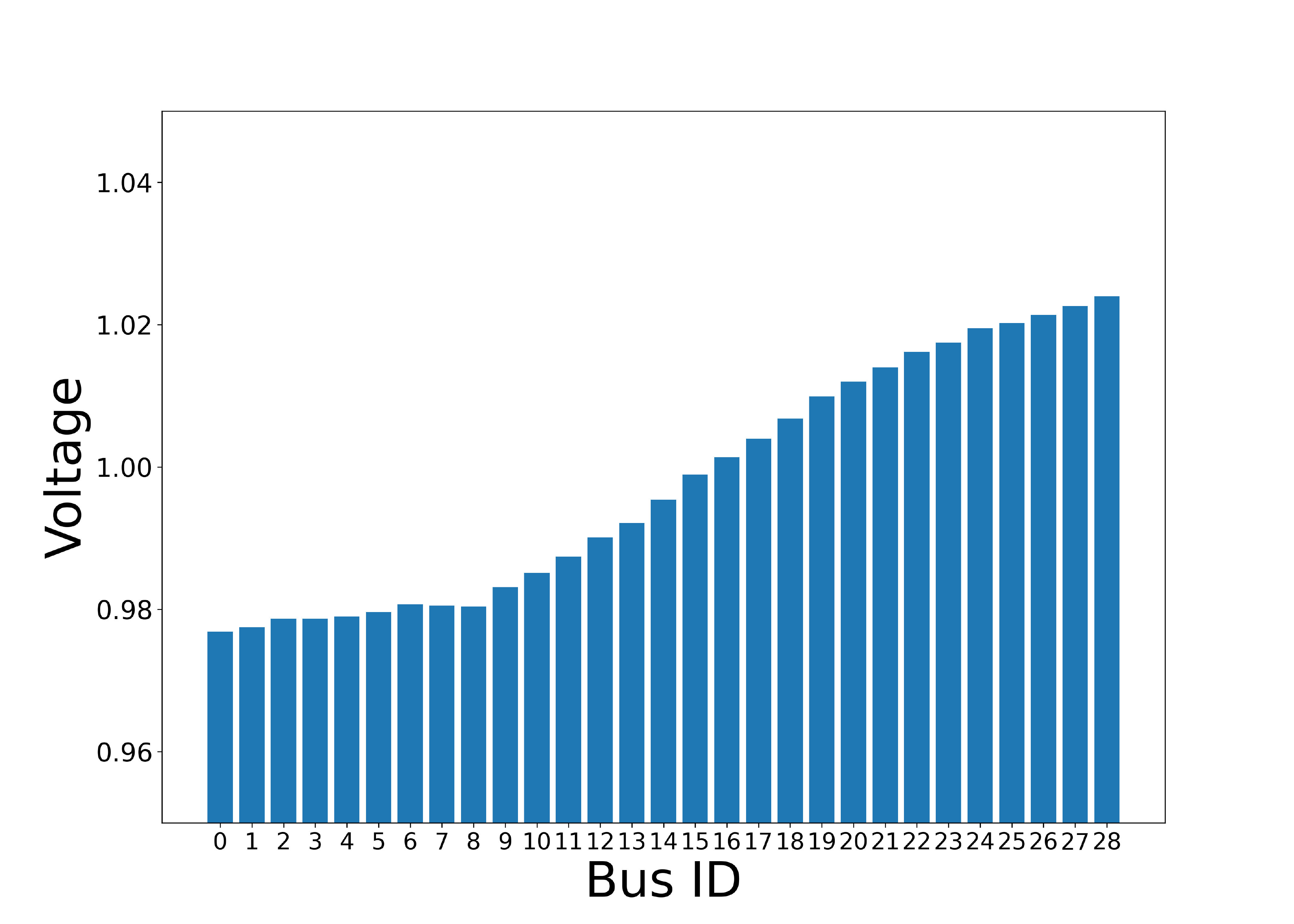}
    }
    \subfigure[Attention Weight]{
        \label{fig:case-attention}
        \includegraphics[scale=0.1]{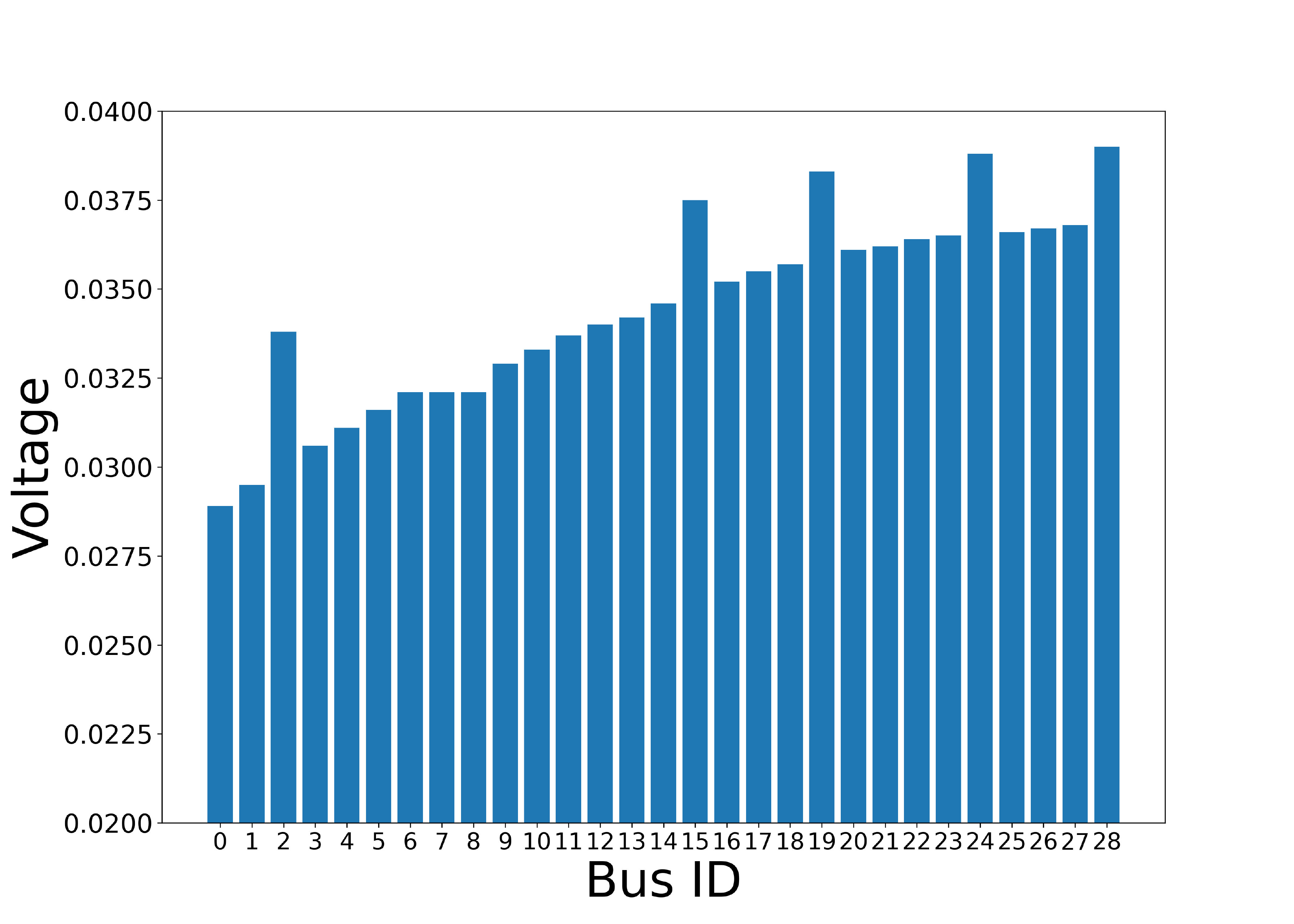}
    }
    \caption{A case study on the 322-bus network. We visualize the attention weights of the special agent in (a). The voltages and attention weights of each nodes are shown in (b) and (c), respectively.}
    \label{fig:case_study}
\end{figure}

As shown in Figure \ref{fig:case_study}, we visualize the attention weights in the embedding aggregation module to figure out which nodes the agent focuses on.
We illustrate a zone in the 322-bus network in Figure \ref{fig:case-network}. Five PVs are installed at different locations in the zone. We select the PV installed in node 15 as the special agent, and record attention weights of the final self-attention layer in the embedding aggregation module.
As shown in Figure \ref{fig:case-voltage}, the voltage of all nodes is within safety range ($0.95 p.u.$ - $1.05 p.u.$).
However, due to the radial topology of the distribution network, nodes at the end of the grid face a greater risk of voltage deviations\cite{OPF}.
The trend of attention weights in Figure \ref{fig:case-attention} shows that the agent pays more attention to the nodes with high voltage based on the topology of this zone.
It is also worthwhile to note that the agent is most concerned with the nodes installed with PVs, which demonstrates that the agent becomes aware of the locations of other PVs in the zone.
This phenomenon verifies the opinion discussed in Section \ref{sec:method} that the T-MAAC captures the nature of the gird and extracts better representations.

\end{document}